\relax 
\documentclass[pra,preprint,amsfonts,showpacs,eqsecnum]{revtex4}         

\usepackage{latexsym}   
\usepackage[]{graphicx} 
\usepackage{times}
\usepackage{bm}

\def\defeq{\mathrel{\stackrel{{\rm def}}{=}}}

\begin{document}   

\title{Framework for quantum modeling of fiber-optical networks: Part~I
\\  (\small Rev. 0.2.5:\ \
  suggestions and corrections welcome)\vadjust{\kern20pt}}

\author{John M. Myers}

\affiliation{Gordon McKay Laboratory, Division of Engineering and Applied
Sciences\\ Harvard University, Cambridge, Massachusetts 02138}

\date{\textbf{31 May 2005}\\[30pt]}

\begin{abstract}

We formulate quantum optics to include frequency dependence in the
modeling of optical networks.  Entangled light pulses available for
quantum cryptography are entangled not only in polarization but also,
whether one wants it or not, in frequency.  We model effects of the
frequency spectrum of faint polarization-entangled light pulses on
detection statistics.  For instance, we show how polarization
entanglement combines with frequency entanglement in the variation of
detection statistics with pulse energy.

Attention is paid not only to single-photon light states but also to
multi-photon states.  These are needed (1) to analyze the dependence
of statistics on energy and (2) to help in calibrating fiber couplers,
lasers and other devices, even when their desired use is for the
generation of single-photon light.

\end{abstract} 

\maketitle
\thispagestyle{empty}

\newpage
\setcounter{page}{1}
\def\thepage{\roman{page}} 

\def\TOCsec#1#2#3{\vspace{0pt}\noindent \hangindent=.5in \noindent \hbox
to .5in{\hfil #1.\ }#2\dotfill #3\vspace{-3.5pt}}
\def\TOCsubsec#1#2#3{\noindent \hangindent=.72in \noindent \hbox to
.5in{}\hbox to .22in{#1.\ \hfil}#2\dotfill #3}
\def\TOCapp#1#2#3{\noindent \hangindent=1.04in \noindent \hbox to
.125in{}Appendix\  #1.\ $\,$#2\dotfill #3}
\def\TOCref#1#2{\noindent \hangindent=.875in \noindent \hbox to
.125in{}#1\dotfill #2}

\centerline{\large\textbf{Contents}}

\vspace{3pt} 

\noindent PART I

\TOCsec{1}{Introduction}{1}

\TOCsubsec{A}{Quantum modeling}{2} 

\TOCsubsec{B}{Aims in developing a framework}{3}

\TOCsubsec{C}{Approach}{4}

\TOCsec{2}{Modes, commutation rules, and light states}{5}

\TOCsubsec{A}{Single-photon state spread over multiple modes}{9}

\TOCsubsec{B}{Single-mode, multi-photon states}{9}

\TOCsubsec{C}{Broad-band coherent states}{10}

\TOCsubsec{D}{General state}{11}

\TOCsubsec{E}{Density matrices and traces}{12}

\TOCsubsec{F}{Partial traces of light states}{13}

\TOCsubsec{G}{Bi-photons: excitation in each of two orthogonal
modes}{15}

\TOCsec{3}{Projections}{16}

\TOCsubsec{A}{Action of single-mode projections on multi-mode
states}{17}

\TOCsubsec{B}{Multi-mode $n$-photon projector}{17}

\TOCsubsec{C}{Number operator}{18}

\TOCsec{4}{Loss and frequency dispersion}{19}

\TOCsubsec{A}{Loss cannot evade ``no cloning''}{19}

\TOCsec{5}{Local quantum fields}{20}

\TOCsubsec{A}{Temporally local hermitian fields}{21}

\TOCsubsec{B}{Time, space, and dispersion}{22}

\TOCsubsec{C}{Projections in terms of local operators}{23}

\TOCsec{6}{Scattering matrix}{23}

\TOCsubsec{A}{Network without frequency mixing}{24}

\TOCsec{7}{Polarized and entangled light states}{25}

\TOCsubsec{A}{Fiber splice (without extraneous modes)}{26}

\TOCsubsec{B}{Coupler}{27}

\TOCsubsec{C}{Entangled states}{27} 

\TOCsubsec{D}{Polarization-entangled states}{28}

\TOCsec{8}{Detection}{29}
  
\TOCsubsec{A}{Simple examples}{30}
 
\TOCsubsec{B}{Model of APD detector for quantum cryptography}{31}

\TOCsubsec{C}{Detection probabilities}{33}

\TOCsubsec{D}{Effect of time bounds on detection}{36}

\TOCsubsec{E}{Detection, energy, and photon subspaces}{36}

\TOCsubsec{F}{Preceding the APD detector by a beam-splitter}{37}

\TOCsec{9}{Polarization-entangled light for QKD}{39}

\TOCsubsec{A}{Bi-photon light states}{39}

\TOCsubsec{B}{Effect of a beam splitter}{41}

\TOCsubsec{C}{Effect of polarization rotation}{41}

\noindent PART II

\TOCsec{10}{Modeling polarization-entangled QKD}{43} 

\TOCsubsec{A}{Outcomes and probabilities}{45}

\TOCsubsec{B}{Light state}{48}

\TOCsubsec{C}{Energy profile}{50}

\TOCsubsec{D}{Calculation of probabilities}{51}

\TOCsubsec{E}{Case I: No frequency entanglement}{56} 

\TOCsubsec{F}{Case II: Limit of extreme frequency entanglement as
$\zeta\to \pm \infty$}{58}

\TOCsubsec{G}{Example numbers}{59}

\TOCapp{A}{Background}{59}

\TOCapp{B}{Operator Lemmas}{60}

\TOCapp{C}{Algebra of frequency-entangled operators}{64}

\TOCapp{D}{Fourier transforms in space and time}{76}

\TOCapp{E}{Expansion of light states in tensor products of 
broad-band\vadjust{\kern-7pt}\hfil\break coherent states}{77}

\TOCapp{F}{MATLAB programs for Section \protect\ref{sec:10}}{78}

\TOCref{References}{97}    

\newpage\clearpage
\setcounter{page}{1}
\def\thepage{\arabic{page}}

\centerline{\large\bf PART I}

\section{Introduction}\label{sec:1}

The complexities of quantum optics, with its multiple integrals over
frequency and wave vectors, tempt one to simplify, and indeed the
groundbreaking equations that launched quantum key distribution (QKD)
were simplified rather drastically, often leaving out altogether the
frequency spectrum of the light involved.  While on one hand the QKD
equations have involved simplifications, on the other hand they invoke
concepts of quantum decision theory, little used in quantum optics,
such as trace distances between density operators as a measure of
their distinguishability.  The motivation for putting the complications
of frequency spectra back into the equations by which we model the
faint light used in QKD comes from recognizing that both in
implementing QKD systems and in designing eavesdropping attacks
against them, frequency spectra play a crucial role.

This report adapts quantum optics to deal as directly as possible with
pulses of weak light propagating though optical fibers.  The equations
introduced here to model faint light give expression to frequency
spectra, including frequency-entanglement; they also define and show
examples of relevant partial traces of density operators for
entangled, frequency-dependent light, needed to make use of an
exceedingly useful relation between entangled-state QKD and QKD
implemented without entanglement.

The report grew from notes on techniques, some borrowed, others
developed from scratch, needed to model a version of BB84 that uses
polarization-entangled light.  Polarization-entangled light from
available sources is also frequency-entangled, and the driving
question was how this frequency entanglement modulates the dependence
of polarization-entangled QKD detection probabilities on mean photon
number.

Some subsequent papers dealing with frequency effects in
polarization-entangled QKD, such as Ref.\ \cite{SPIE05}, use the
techniques and results of this report, and, in particular use certain
convolution integrals that are described in Sec.~\ref{sec:10} and
investigated in detail in Appendix~\ref{app:C}, with accompanying MATLAB
programs given in Appendix~\ref{app:F}.

\subsection{Quantum modeling}

By definition, quantum modeling invokes equations constrained in form
to those of quantum mechanics, expressing a joint probability
distribution of (theoretical) outcomes in terms of an initial density
operator $\rho$ at time $t =0$, a hamiltonian evolution operator $H$,
and a resolution of the identity consisting of a set of non-negative
operators $M_j$ satisfying $\sum_j M_j \le 1$.  These engender a
probability of a (theoretical) outcome $j$ \cite{0404113}: 
\begin{equation} \Pr(j) =
\mbox{Tr}[e^{-itH/\hbar}\rho e^{itH/\hbar}M_j].  \label{eq:form}
\end{equation} 
Whether or not one makes it explicit, at $\rho$, $H$, and $M_j$ are
functions of parameters that one views as under experimental
control; all that a model can say is said in terms of how the
probabilities depend, via $\rho$, $H$, and $M_j$, on these parameters.
A system of equations for modeling particular devices includes
equations that specify properties of the $H$, $\rho$, and $M_j$, thus
specializing the probability distribution $\Pr(j)$.  The outcome
$j$ can be a list of components, {\em e.g.}\ one component for
each of several detectors, in which case $\Pr(j)$ is viewed as a joint
probability for the components of the outcome. 

In choosing quantum equations to model experiments with light,
one expresses light by one or another density operator $\rho$, and one
expresses detecting devices by operators $M_j$, possibly augmented by
probe particles, as discussed in \cite{0404113}.  As is well known,
the boundary between preparation of light and its detection is chosen
by the modeler, and can be pushed around \cite{vN}.  Different choices
of $\rho$, $H$, or $M_j$ set up different quantum models.  Implicitly
or explicitly, $\rho$ is a function of variables that express
the setting of various knobs on the laser and/or other devices that
generate the light, and $M_j$ is a function of variables that
express knob settings on the detectors, such as those that control
polarizing filters.  To claim that a set of equations of quantum
mechanics describes an arrangement of devices is to claim that the
probabilities calculated from the equations more or less fit relative
frequencies of experimental outcomes obtained, for some range of knob
settings, from the arrangement of devices.

As discussed in \cite{0404113}, choosing equations to model an
experiment takes guesswork, indeed, two layers of guesswork.  From
experiments with devices and a first layer of guesswork \cite{ams,JOptB},
one abstracts experimental relative frequencies that can be compared
with modeled probabilities.  Demanding an approximate fit to
experimental relative frequencies puts a constraint on probabilities
of outcomes as functions of knob settings, and hence establishes a
property of states and operators that can be judged as fitting an
experimental situation.  Still, diverse systems of equations can always
be found to agree with any given set of experimental relative
frequencies and yet disagree among themselves in probabilities that
they generate for arrangements of devices not yet explored.  For this
reason, arriving at equations of quantum mechanics by which to
describe the measured behavior of QKD devices requires reaching beyond
logic to make what may as well be called a guess.  For this reason,
the sensible use of equations in modeling QKD is hardly to `prove
security' but instead to help in achieving transmission of keys in the
face of practical obstacles and to design eavesdropping attacks.

By recognizing an irreducible freedom in choosing systems of equations
to describe an arrangement of devices, we can clarify the mathematics
of modeling, unencumbered by physical interpretations that are always
subject to choice.  Besides helping with QKD, the techniques of
modeling presented here can also serve other uses of faint light that
generate interesting joint detection statistics expressible in quantum
mechanics but unknown to classical physics, correlations that allow
the invention of new kinds of ``cameras'' with which to see and
respond to the physical world.

\subsection{Aims in developing a framework}
\begin{figure}[t]  
\begin{center}
\includegraphics[width=4.6in]{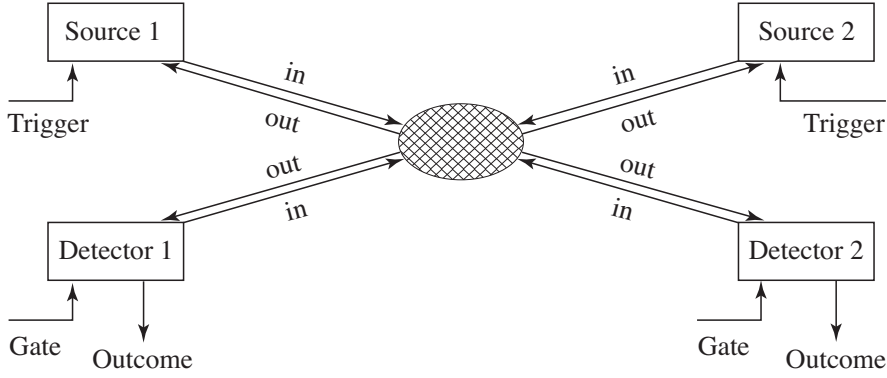}
\end{center}
\vspace{-24pt}
\caption{Optical network.}\label{fig:1}
\end{figure}

Picture an experimental network involving light sources and detectors
linked by fibers as shown in Fig.\ \ref{fig:1}, where the blob in the
middle can include phase shifters, fiber couplers, and attenuators, as
well as conversions from fiber to free space and back again to fiber.
Mainly I discuss so-called single-mode fibers; (most of these actually
support two polarizations).  The various fibers of a network need not
be alike; for instance they can vary in their propagation constants
and in their attenuation.  Although most of the discussion is in terms
of fiber, free-space links can also be included.  The mathematical
framework offered allows modeling the variety of responses encountered
experimentally in light detection \cite{0404113}; in particular, the
response to single-photon states need not be binary.

The aim in developing this framework for the analysis of optical networks
is this:
\begin{enumerate}
\item
Provide equations to express
violations of Bell inequalities.

\item
Provide for modeling networks assembled from smaller pieces, like
tinker toys, by splicing fibers of one to fibers of another.

\item
Provide for convenient expression of pieces of networks
that are free of frequency conversion but that have diverse fibers
with diverse propagation constants at any single frequency.  I.e
provide mathematics convenient for expressing single-frequency
modes for networks (or parts of networks) containing fibers that
differ in their propagation constants.\vadjust{\kern-9pt}
\end{enumerate}

\subsection{Approach}
\begin{figure}[t] 
\begin{center}
\includegraphics[width=3.25in]{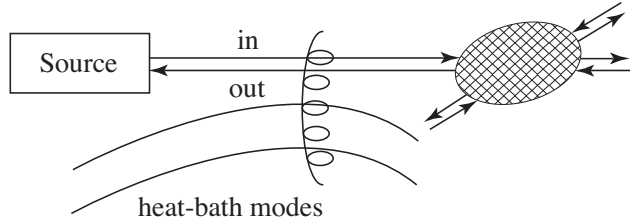} 
\end{center}
\vspace{-15pt}
\caption{Dissipation modeled by coupling of desired modes to loss
modes.\vadjust{\kern-6pt}}\label{fig:2}
\end{figure}

We split up the task of modeling a quantum network into the following
modules:
\begin{enumerate}
\item Develop mathematics for the quantum mechanics of a set of uncoupled,
lossless transmission lines, each line expressed by a set of modes, where
each mode supports a range of frequencies propagating in two
directions, denoted ``$+$'' and ``$-$''.
\begin{enumerate}
\item Corresponding to each mode, introduce a creation operator, and
define single-mode quantum states as the vacuum state acted on by
superpositions of these operators over some frequency band.
\item Extend to many modes.

\item Define projection operators, each of which corresponds to a subspace
of $n$ photon states on a Hilbert space that is a tensor product of
some number of modes.
\begin{enumerate}
\item Use lattice of projections to define subspaces. 
\end{enumerate}
\item Define detection operators on such states.

\item Perform calculations by use of commutation relations.
\end{enumerate}
\item Use scattering theory to approximate interacting transmission lines
by operators that convert an in-state on non-interacting lines to an
out-state on non-interacting lines. (See Fig.\ \ref{fig:1}.)
 
\item Model loss and dispersion by introducing coupling of desired modes
to extraneous modes, as illustrated in Fig.\ \ref{fig:2}.

\item Model frequency preserving interactions among transmission lines
by unitary transformations acting on creation operators for all
desired and extraneous modes. 

\item Sketch two applications to a network for quantum key distribution,
in which these concepts and techniques work together.\vadjust{\kern-10pt}
\end{enumerate}

\section{Modes, commutation rules, and light states}\label{sec:2}

In analogy with classical electromagnetics, we will model a path, such
as an optical fiber, as a system of modes, with each mode supporting a
continuous range of frequencies.  We assume a vacuum state
$|0\rangle$, normalized so that
\begin{equation}\langle 0 | 0 \rangle = 1. 
\label{eq:norm0}
\end{equation}
Other states are defined by various creation operators acting on the
vacuum state.  These in turn are defined in terms of more singular
single-frequency creation operators, as follows. For modes $a$, $b$,
\dots, one could introduce annihilation operators $\hat{a}$, $\hat{b}$,
\dots; however we avoid the clutter of the ``hats,'' so that, for
example, $b$ expresses a mode name in some contexts but in other contexts
expresses an annihilation operator for that mode.  Each mode is
bi-directional; we call one direction ``$+$'' and the other ``$-$''. Let
$a_+^\dag(\omega)$ be the creation operator for excitation at an angular
frequency $\omega$ in the ``$+$'' direction, and let $a_-^\dag(\omega)$
be the creation operator for propagation in the ``$-$'' direction.  The
operators $a_+$ and $a_-$ are defined for non-negative $\omega$.  It is
often convenient to combine the ``$+$'' and ``$-$'' operators into a
single operator.  Since these are both defined for $\omega \ge 0$, we
can define $a(\omega)$ for $-\infty < \omega < \infty$ by
\begin{equation}
a(\omega)= \left\{ \begin{array}{ll} a_+(\omega)\quad&\mbox{
    for }\omega > 0, \\
a_-(-\omega)&\mbox{ for }\omega <  0.
\end{array}\right. 
\label{eq:adef}
\end{equation}  

For any given mode, a single-photon state is any state of the
form
\begin{equation}a_f^\dag|0\rangle, \end{equation}
where the creation operator $a_f^\dag$ is defined by 
\begin{equation} 
a_f^\dag \defeq \int_{-\infty}^\infty d\omega\,
f(\omega)a^\dag(\omega),
\label{eq:afdef}\end{equation} 
and $f$ is any square-integrable complex-valued function of $\omega$,
normalized so that 
\begin{equation}
\int d\omega\, |f(\omega)|^2 = 1,
\label{eq:normf}
\end{equation}
where in this and the following integrals, the integration limits are
$-\infty$ and $\infty$ unless otherwise specified; the role of
negative frequencies will be discussed shortly.  For the given mode,
$a^\dag(\omega)$ is the creation operator for excitation at an angular
frequency $\omega$. It is the adjoint of an annihilation operator
$a(\omega)$, and the commutation relation between the two is
\begin{eqnarray} [a(\omega), a^\dag(\omega')] &  = & 
\delta(\omega - \omega'), \label{eq:comm0}
 \\ {} [a(\omega), a(\omega')] & = & 0.
\label{eq:comm1}\end{eqnarray}
This relation is a simplification appropriate to fiber modes of the
commutation relation in \cite{yuen}.  The annihilation operator is
required to satisfy the rule
\begin{equation} (\forall\ \omega)\ \ a(\omega)|0\rangle = 0,
\end{equation} which from the adjoint of Eq.\ (\ref{eq:afdef}) implies
that
\begin{equation}
a_f|0\rangle = 0 .
\label{eq:kill}\end{equation}
{}From Eq.\ (\ref{eq:comm0}) follows the commutation rule
for any functions $f$ and $g$:
\begin{eqnarray} [a_g,a_f^\dag] & = & \int\!\! \int d\omega\,  
d\omega'\, 
g^*(\omega)f(\omega')[a(\omega),a^\dag(\omega')]\nonumber\\ & = &
\int\!\! \int d\omega\, d\omega'\, g^*(\omega)f(\omega')\delta(\omega -
\omega') \nonumber \\ & = & \int d\omega\, g^*(\omega)f(\omega),
\label{eq:commf}\end{eqnarray}
where the asterisk denotes the complex conjugate.  This
can be written more compactly as
\begin{equation}
[a_g,a_f^\dag] = (g,f),
\label{eq:commAbr}
\end{equation}
where we define the inner product of functions
\begin{equation}
(g,f) =   \int d\omega\, g^*(\omega)f(\omega).
\label{eq:inProd}
\end{equation}
{}From this with normalized functions $g=f$, it follows that $(f,f) = 1$
by Eq.\ (\ref{eq:normf}), and we see that $a_f^\dag$ and $a_f$ are boson
creation and annihilation operators satisfying $[a_f,a_f^\dag]=1$.  As
a result, we can calculate the norm of a single-photon state
$a_f^\dag|0\rangle$ to be
\begin{eqnarray}\| a_f^\dag|0\rangle \| & = & (\langle 0| a_f a_f^\dag
|0\rangle)^{1/2} \nonumber \\ & = & [\langle 0|(a_f^\dag a_f +
  1)|0\rangle]^{1/2}  = \langle 0|0\rangle^{1/2} = 1,
\end{eqnarray}
where the next-to-last equality follows from Eq.\ (\ref{eq:kill}).
For probabilities to make sense, we must require a finite inner
product on quantum states, which requires normalizable states.  This
makes some bandwidth necessary: there can be no normalizable states at
a single frequency.  Because $a^\dag(\omega)|0\rangle$ has no norm, we
call $a^\dag(\omega)$ an \textit{improper} operator.

The hamiltonian operator for the mode (without zero-point energy) is
\begin{equation} H = \hbar \int d\omega \, |\omega| 
 a^\dag(\omega)a(\omega). \label{eq:energy1}\end{equation}

\vspace{3pt}

\noindent Example 1: The expectation energy of a single-photon state
$a^\dag_f|0\rangle$ defined by the energy operator of Eq.\
(\ref{eq:energy1}) is:
\begin{eqnarray} \langle 0 |a_f H a_f^\dag |0 \rangle & = & \hbar
  \langle 0 | \int\!\! \int\!\! \int d\omega_1\, d\omega_2\, d\omega_3\,
f^*(\omega_1)a(\omega_1) |\omega_2|
a^\dag(\omega_2)a(\omega_2)f(\omega_3)a^\dag(\omega_3)|0 \rangle
\nonumber \\ & = & \hbar \int d\omega \, |\omega|\, |f(\omega)|^2,
\end{eqnarray} where the second equality is obtained using Eq.\
(\ref{eq:comm0}) and integrating out the $\delta$-functions.
Thus if $f(\omega)$ is concentrated around some $\omega_0$, one
finds by this prescription a photon energy of about $\hbar
|\omega_0|$.\looseness=-1

\vspace{3pt}

Commutation rules enable the calculation of probabilities of
the form of Eq.\ (\ref{eq:form}).  For instance, for a pure
state, the right-hand side of Eq.\ (\ref{eq:form}) takes the form
\begin{equation}
\langle 0 |\int d\omega_1 \cdots d\omega_n\, g(\omega_1,\ldots,
\omega_n) \mbox{Pol}(a_i(\omega_j),a_i^\dag(\omega_j),b_\ell(\omega_k),
b_\ell^\dag(\omega_k),\ldots)|0 \rangle, 
\end{equation} 
where $\mbox{Pol}$ is a polynomial in creation and annihilation
operators subject to $\delta$-function commutation relations, and
every annihilation operator acting from the left on the vacuum state
$|0\rangle$ gives 0, as does every creation operator acting on the
right of $\langle 0|$.  The standard method of evaluating such a
probability is to use the commutation relations to put $\mbox{Pol}$ in
normal order.  In the cases of interest, every power of every creation
operator is paired with the same power of the corresponding
annihilation operator, so that the only terms that contribute after
the commutations have the form
\begin{eqnarray}
\lefteqn{\langle 0 |\int d\omega_1 \cdots d\omega_n\,
g(\omega_1,\ldots,
\omega_n) Q(\omega_1,\ldots, \omega_n)|0 \rangle}\quad\nonumber\\
& = & \int d\omega_1
\cdots d\omega_n\, g(\omega_1,\ldots, \omega_n) Q(\omega_1,\ldots,
\omega_n), 
\end{eqnarray} 
where $Q$ is a sum of products of
$\delta$-functions, and the equality follows from Eq.\ (\ref{eq:norm0}).

\subsection{Single-photon state spread over multiple modes}
Two modes $a$ and $b$ are called orthogonal if and only if
\hbox{$[a,b^\dag] = [a,b] = 0$}.  Superpositions of single-photon states
across orthogonal modes $a$ and $b$ have the form 
\begin{equation}
(c_1 a_f + c_2 b_g)^\dag|0\rangle,
\label{eq:mulmode}
\end{equation} where $\sum_j |c_j|^2 = 1$.

\subsection{Single-mode, multi-photon states}\label{subsec:2B}
To construct the most general single-mode $n$-photon states, we
introduce notation for multi-photon operators in which there are as
many frequency variables as there are operator factors:
\begin{equation} 
h\!:\!a^{\dag n}  \defeq   \int\!\cdots\!\int
  d\omega_1\cdots d\omega_n\, h(\omega_1,\ldots, \omega_n)
  a^\dag(\omega_1)\cdots a^\dag(\omega_n).
  \label{eq:nphot}\end{equation}
Because the creation operators commute with one another, all that
matters about $h$ is the part of it symmetric under interchange of
arguments, denoted
\begin{equation}
\mathcal{S}(\omega_1,\ldots,\omega_n)\, h(\omega_1,\ldots, \omega_n)
  \defeq  \frac{1}{n!}\,\sum_{\pi \in S_n}\,
  h(\omega_{\pi 1},\ldots, \omega_{\pi n}),
\label{eq:Sdef}
\end{equation} 
where $S_n$ denotes the permutation group of order $n$.
Thus we have
\begin{equation}
h\!:\!a^{\dag n} = \mathcal{S}(\omega_1,\ldots,\omega_n)\,h\!:\!a^{\dag
n}.\label{eq:sym1}
\end{equation}

The normalization condition
\begin{equation}
 \int\!\cdots\!\int d\omega_1 \cdots d\omega_n\,
  |\mathcal{S}(\omega_1,\ldots,\omega_n)\, h(\omega_1,\ldots,
  \omega_n)|^2 = 1
\label{eq:nnorm}
\end{equation} 
assures unit norm for the single-mode $n$-photon state
\begin{equation}
\frac{1}{\sqrt{n!}}\,(h\!:\!a^{\dag n})|0\rangle .
\label{eq:nphstate}
\end{equation} 
The adjoint works according to 
\begin{equation}[(h\!:\!a^{\dag n})|0\rangle]^\dag = \langle
0|(h^*\!:\!a^n).
\end{equation}

\subsubsection{Inner product of two single-mode, multi-photon states}
Consider two modes that need not be orthogonal, such as two linearly
polarized modes $a$ and $b$ with an angle $\theta$ between them, so
the commutation relation between them is $[b(\omega),a^\dag(\omega')]
= \cos \theta\,\delta(\omega-\omega')$.  For symmetric functions
$h$ of $n$ arguments and $h'$ of $m$ arguments, the 
usual manipulations show that
\begin{eqnarray}
\frac{1}{\sqrt{n!m!}}\,  \langle 0|(h'^*\!:\!b^m)(h\!:\!a^{\dag
n})|0\rangle & = & (\mathcal{S}(\omega_1,\ldots,\omega_m)\,h',
\mathcal{S}(\omega_1,\ldots,\omega_n)\,h) \cos^n \theta ,
\end{eqnarray}
where we define the inner product of the multi-variable
functions $h$ and $h'$ as
\begin{equation} 
(h',h) = \left\{\begin{array}{l}0,\quad\mbox{if $h$ and $h'$ are unlike in
    number of arguments,} \\ \int\!\cdots\!\int d\omega_1 \cdots
d\omega_n\, h'_n{\!}^*(\omega_1,\ldots, \omega_n)h_n(\omega_1,\ldots,
\omega_n),\quad\mbox{otherwise}.
\end{array}\right.
\end{equation} 
Thus for $h = h'$ and $h$ normalized per Eq.\ (\ref{eq:nnorm}), the
inner product of the two $n$-photon states is just $\cos^n \theta$.

\subsubsection{Energy of single-mode, $n$-photon state}
For  $h$ normalized per Eq.\ (\ref{eq:nnorm}) and the energy
operator $H$ of Eq.\ (\ref{eq:energy1}), the expectation
energy for the state $n!^{-1/2}(h\!:\!a^{\dag n})|0\rangle$ is
\begin{equation} 
\frac{1}{n!} \langle 0|(h^*\!:\!a^n)H (h\!:\!a^{\dag n})| 0\rangle = n \hbar
\omega_{h},
\end{equation}
where we define an average angular frequency
\begin{equation}\omega_{h} \defeq 
\int\!\cdots\!\int d\omega_1\cdots d\omega_n\,|\omega_1|\,
|\mathcal{S}(\omega_1,\ldots,\omega_n)\,h(\omega_1,\ldots,\omega_n)|^2.
\label{eq:centerF}
\end{equation}

\subsection{Broad-band coherent states}\label{subsec:2C}
As a special case of a two-photon state, one can choose any normalized
function $f$ and define $h(\omega_1,\omega_2)= 2^{-1/2}
f(\omega_1)f(\omega_2)$ to produce a two-photon state $2^{-1/2}
(a_f^\dag)^2|0\rangle$.  This immediately generalizes to higher powers
of $a_f^\dag$, leading to a `broad-band' coherent state. From Eq.\
(\ref{eq:commf}) we see that $a_f^\dag$ and $a_f$ are boson creation
and annihilation operators satisfying $[a_f,a_f^\dag]=1$, so that
Louisell's discussion of coherent states [\onlinecite{louisell}, 
Sec.\ 3.2] applies to them.  Hence we can define $n$-photon $f$-states by
powers of $a_f^\dag$ acting on the vacuum state $n$ times, producing
the normalized state $(n!)^{-1/2}(a_f^\dag)^n|0\rangle$.  Coherent
$f$-states can be defined as super\-positions of these states, just as
in Louisell [\onlinecite{louisell}, p.~104]:  
\begin{eqnarray} |\alpha,a_f\rangle &\defeq&
\exp\left(-\frac{1}{2}\,|\alpha|^2\right)\sum_{n=0}^\infty\,
(n!)^{-1}(\alpha a_f^\dag)^n|0\rangle \nonumber \\
& = & \exp\left(-\frac{1}{2}\,|\alpha|^2\right)\exp(\alpha a_f^\dag)
|0\rangle.
\label{eq:coh}
\end{eqnarray}
This is an example of a calculation that proceeds exactly as if
frequency dependence were collapsed, so that $a_f$ works like a 
simple oscillator annihilation operator.  In contrast, frequency
dependence matters in the commutation rule
\begin{equation} [a_g, a_f^\dag] = \int d\omega\,
    g^*(\omega)f(\omega),
\end{equation} 
which follows from Eqs.\ (\ref{eq:comm0}) and (\ref{eq:comm1}).

To calculate the energy of the coherent state, note first that the
commutation rules Eqs.\ (\ref{eq:comm0}) and (\ref{eq:comm1}) imply for
the coherent state defined by Eq.\ (\ref{eq:coh})
\begin{equation} a(\omega)|\alpha, a_f\rangle = \alpha
  f(\omega)|\alpha, a_f\rangle.
\label{eq:b5}
\end{equation}
{}From this and Eqs.\ (\ref{eq:energy1}) and (\ref{eq:normf})
follows the expectation energy
\begin{equation}
\langle\alpha,a_f|H|\alpha,a_f\rangle  =   
\hbar \omega_f |\alpha|^2 ,
\label{eq:cohenergy}
\end{equation}
where we have defined a mean absolute angular frequency 
\begin{equation}
\omega_f \defeq  \int d\omega\,|\omega|\,|f(\omega)|^2.
\end{equation}

\subsection{General state}
The general state is a sum of terms, not necessarily normalized, each
of the form
\begin{equation}
|\psi\rangle = f\!:\!\prod_{j=1}^Ja_j^{\dag n_j}|0\rangle.
\label{eq:formTerm}
\end{equation}
Symmetries under interchange of variables are best expressed in a notation
intermediate between that of writing out the integrals and the compact
``colon'' notation.  We expand the shorthand $f\!:\!a^{\dag n}$ to
$f(\bm{\omega})\!:\!a^{\dag n}(\bm{\omega})$, understanding that with
an exponent $n$ involved, $(\bm{\omega})$ is short for a list of $n$
frequency variables and $a^{\dag n} (\bm{\omega})$ is short for
$a^\dag(\omega_1)\cdots a^\dag(\omega_n)$. 
Written in this notation, Eq.\ (\ref{eq:formTerm}) becomes
\begin{equation}
|\psi\rangle =
  f(\bm{\omega}_1,\ldots,\bm{\omega}_J)\!:\!\prod_{j=1}^Ja_j^{\dag
  n_j}(\bm{\omega}_j)|0\rangle;
\end{equation} 
where $(\bm{\omega}_j)$ is a list of $n_j$ frequency variables.
Symmetry under interchange
of creation operators implies, as in Eq.\ (\ref{eq:sym1}),
\begin{equation}
f(\bm{\omega})\!:\!a^{\dag n}(\bm{\omega}) = 
[\mathcal{S}(\bm{\omega})f(\bm{\omega})]\!:\!a^{\dag
  n}(\bm{\omega}),
\end{equation}
where the symmetry operator $\mathcal{S}$ is defined in Eq.\
(\ref{eq:Sdef}); this generalizes to
\begin{equation}
f(\bm{\omega}_1,\ldots,\bm{\omega}_J)\!:\!\prod_{j=1}^Ja_j^{\dag
  n_j}(\bm{\omega}_j)|0\rangle =
  \left[\mathcal{S}(\bm{\omega}_1)\cdots
  \mathcal{S}(\bm{\omega}_J) f(\bm{\omega}_1,
  \ldots,\bm{\omega}_J)\right]\!:\!\prod_{j=1}^Ja_j^{\dag
  n_j}(\bm{\omega}_j)|0\rangle.
\end{equation}
Sometimes we write such expressions a little more compactly,
using the convention that $\bm{\omega}_n =
(\bm{\omega}_1,\ldots,\bm{\omega}_J)$ as 
\begin{equation}
f(\bm{\omega}_n)\!:\!\prod_{j=1}^Ja_j^{\dag
  n_j}(\bm{\omega}_j)|0\rangle =
  \left[\left(\prod_{j=1}^J\mathcal{S}(\bm{\omega}_j)\right)
  f(\bm{\omega}_n)\right]\!:\!\prod_{j=1}^Ja_j^{\dag
  n_j}(\bm{\omega}_j)|0\rangle.
\end{equation}
We can distinguish one set of modes from another by replacing some of
the $a_j$ by other labels, such as $b_j$.  For example, $a$-modes can
refer to Alice while $b$-modes refer to Bob.  This leads to a general
term in a state expansion of the form
\begin{equation}
|\psi\rangle =
  f(\bm{\omega}_n,\tilde{\bm{\omega}}_n)\!:\!\left(\prod_{j=1}^Ja_j^{\dag
  n_j}(\bm{\omega}_j)\right)\left(\prod_{k=1}^Kb_k^{\dag
  m_k}(\tilde{\bm{\omega}}_k)\right)|0\rangle.
\end{equation}

\subsection{Density matrices and traces}
A density matrix is a sum 
\begin{equation}
\rho = \sum_n w_n|\Psi(n)\rangle \langle \Psi(n)|,
\end{equation} 
where the $|\Psi(n)\rangle$ are unit vectors, $w_n \ge 0$, and $\sum_n
w_n = 1$; or an integral
\begin{equation}
\rho = \int\!\! w(u)\,du \,|\Psi(u)\rangle\langle\Psi(u)|,
\end{equation} 
with the $|\Psi(u)\rangle$ unit vectors, $w(u) \ge 0$ and $\int\! du\,
w(u) = 1$. More generally, a density matrix can be any convex sum of
these discrete and continuous types.
 
We need the trace of a density operator multiplied by a bounded
operator $M$.  Although in the context of infinite-dimensional spaces
the trace is sometimes defined only for positive operators
\cite{sewell,ak}, we want to apply a trace to terms that occur when a
pure-state density operator is expanded (i.e., $|\Psi\rangle
\langle \Psi| = \sum_{m,n}c_m c_n^* |\Psi_m\rangle \langle
\Psi_n|$. For this we define
\begin{equation} 
\mbox{Tr}[(|\psi\rangle\langle\phi|)M] =
\langle\phi|M|\psi\rangle,
\end{equation}
and define more general traces by linearity.  We then find
\begin{eqnarray}
\mbox{Tr}(\rho M)& \stackrel{\rm def}{=} & \sum_n w_n \langle \Psi(n)|
  M|\Psi(n)\rangle \mbox{ (discrete case)},\nonumber \\
\mbox{Tr}(\rho M)& \stackrel{\rm def}{=} & \int\!\! du \,w(u)\langle
\Psi(u)|
  M|\Psi(u)\rangle \mbox{ (continuous case)}.
\end{eqnarray}

\subsection{Partial traces of light states}
Probabilities for detection of light states often involve {\em partial
traces}, defined for an operator on a tensor-product space
$\mathcal{H}_A \otimes \mathcal{H}_B$ by linearity from the following
special case.  For $|a\rangle, |a'\rangle \in \mathcal{H}_A$ and
$|b\rangle, |b'\rangle \in \mathcal{H}_B$,
\begin{eqnarray}
\mbox{Tr}_B(|a\rangle|b\rangle\langle b'|\langle a'|)&  = &
(\langle b'|b\rangle)|a\rangle\langle a'|, \nonumber \\
\mbox{Tr}_A(|a\rangle|b\rangle\langle b'|\langle a'|)&  = &
(\langle a'|a\rangle)|b\rangle\langle b'|.
\label{eq:traceRule}
\end{eqnarray}
{}From this it follows that for any bounded operator $M_A$ that
acts only on $\mathcal{H}_A$, we have
\begin{equation}
\mbox{Tr}_A[(|a\rangle|b\rangle\langle b'|\langle a'|)M_A] =
(\langle a'|M_A|a\rangle)|b\rangle\langle b'|.
\label{eq:Aop}
\end{equation}
Here is an example.  If a detection operator is of the form
$M_{A,j}\otimes 1_B$, then
\begin{equation}
\mbox{Tr}_{AB}[\rho_{AB}(M_{A,j}\otimes 1_B)]  =
\mbox{Tr}_A(\rho' M_{A,j}),
\end{equation}
\begin{equation}
\rho' = \mbox{Tr}_B(\rho_{AB}).
\end{equation} 
The partial trace over $b$-modes of any light state $\rho$ is
calculated from the commutation relations applied to the inner product
of $b$-mode factors in the usual way, supported by the notation
developed in the preceding subsection.
For example, one can deal with the case of a
single $a$-mode and a single $b$-mode as follows.  Suppose we have
\begin{equation} |\psi_{mn}\rangle = \int\!\!d\bm{\omega}\,
  d\tilde{\bm{\omega}}\,g(\bm{\omega},\tilde{\bm{\omega}})
a^{\dag m}(\bm{\omega})b^{\dag n}(\tilde{\bm{\omega}})|0\rangle,
\end{equation}
where we abbreviate $\bm{\omega}= \omega_1,\ldots,\omega_m$,
$\,\tilde{\bm{\omega}} = \tilde{\omega}_1,\ldots,\tilde{\omega}_n$,
$\,a^{\dag m} (\bm{\omega}) = a^\dag(\omega_1)\cdots
a^\dag(\omega_m)$, etc.  Similarly,
suppose\vadjust{\kern-6pt}
\begin{equation} |\phi_{m'n'}\rangle = \int\!\! d\bm{\omega}'
  d\tilde{\bm{\omega}}'
h(\bm{\omega}',\tilde{\bm{\omega}}') a^{\dag
m'}(\bm{\omega}')b^{\dag n'}(\tilde{\bm{\omega}}')|0\rangle.
\end{equation}
Then, for an operator $M_j(a)$ that commutes with
$b(\tilde{\bm{\omega}})$ and $b^\dag(\tilde{\bm{\omega}})$,
\begin{equation}
\mbox{Tr}_{ab}[(|\psi_{mn}\rangle\langle\phi_{m'n'}|)M_j(a)]
= \mbox{Tr}_a[\rho'M_j(a)],\vadjust{\kern-6pt}
\end{equation}
where
\begin{eqnarray}
\rho' & = & \mbox{Tr}_b(|\psi_{mn}\rangle\langle\phi_{m'n'}|)
\nonumber \\ & = & \int\!\!d\bm{\omega} d\bm{\omega}' \left[\int\!\!
d\tilde{\bm{\omega}}d\tilde{\bm{\omega}}'
g(\bm{\omega},\tilde{\bm{\omega}})
h^*(\bm{\omega}', \tilde{\bm{\omega}})\langle
0_b|b^{n'}(\tilde{\bm{\omega}}')b^{\dag
n}(\tilde{\bm{\omega}})|0_b\rangle \right]a^{\dag
m}(\bm{\omega})|0_a\rangle \langle0_a|a^{m'}(\bm{\omega}') \nonumber
\\ & = & n! \delta_{nn'}\int\!\! d\bm{\omega} d\bm{\omega}'
\left[\int\!\!  d\tilde{\bm{\omega}}\,g(\bm{\omega},\tilde{\bm{\omega}})
\mathcal{S}(\tilde{\bm{\omega}})h^*(\bm{\omega}',
\tilde{\bm{\omega}})\right]a^{\dag m}(\bm{\omega})|0\rangle
\langle0|a^{m'}(\bm{\omega}'),
\end{eqnarray}
where the second equality follows from Lemma (B16) of Appendix~\ref{app:B}
applied to the $b$-modes.  Similarly, one calculates for the trace over
the
$a$-modes
\begin{eqnarray}
\lefteqn{\mbox{Tr}_a(|\psi_{mn}\rangle\langle\phi_{m'n'}|)}
\nonumber \\ & = & \int\!\!d\tilde{\bm{\omega}} d\tilde{\bm{\omega}}'
\left[\int\!\!  d\tilde{\bm{\omega}}d\tilde{\bm{\omega}}'
g(\bm{\omega},\tilde{\bm{\omega}}) h^*(\bm{\omega}',
\tilde{\bm{\omega}})\langle 0_a|a^{m'}(\bm{\omega}')a^{\dag
m}(\bm{\omega})|0_a\rangle \right]b^{\dag
n}(\tilde{\bm{\omega}})|0_b\rangle
\langle0_b|b^{n'}(\tilde{\bm{\omega}}') \nonumber \\ & = & m!
\delta_{mm'}\int\!\! d\tilde{\bm{\omega}} d\tilde{\bm{\omega}}'
\left[\int\!\!  d\bm{\omega}g(\bm{\omega},\tilde{\bm{\omega}})
\mathcal{S}(\bm{\omega})h^*(\bm{\omega},
\tilde{\bm{\omega}}')\right]b^{\dag n}(\tilde{\bm{\omega}})|0\rangle
\langle0|b^{n'}(\tilde{\bm{\omega}}').
\end{eqnarray}

The more general case involves complications, but here it is.
Let $n = \sum_{j=1}^Jn_j$ and $m= \sum_{k=1}^Km_k$ and similarly
for primed quantities; let
\begin{eqnarray}
|\psi\rangle & = &
 h(\bm{\omega}_n,\tilde{\bm{\omega}}_m)\!:\!\left(\prod_{j=1}^J
 a_j^{\dag n_j}(\bm{\omega}_j)\right)\left(\prod_{k=1}^Kb_k^{\dag
 m_k}(\tilde{\bm{\omega}}_k)\right)|0\rangle, \nonumber \\
 |\psi'\rangle & = &
 h'(\bm{\omega}'_{n'},\tilde{\bm{\omega}}'_{m'})\!:\!
 \left(\prod_{j=1}^{J'} a_j^{\dag
 n'_j}(\bm{\omega}'_j)\right)\left(\prod_{k=1}^{K'}b_k^{\dag
 m'_k}(\tilde{\bm{\omega}}'_k)\right)|0\rangle.
\end{eqnarray}
On expanding the `colon' notation and abbreviating using
$d\bm{\omega}_n$ for $d\bm{\omega}_1d\bm{\omega}_2\cdots
d\bm{\omega}_J$ and $d\tilde{\bm{\omega}}_m$ for
$d\tilde{\bm{\omega}}_1d\tilde{\bm{\omega}}_2\cdots
d\tilde{\bm{\omega}}_K$, one finds
\begin{eqnarray}\lefteqn{
\mbox{Tr}_a(|\psi\rangle\langle\psi'|)} \nonumber \\ &=&
\int\!\!d\tilde{\bm{\omega}}_m d\tilde{\bm{\omega}}'_{m'}\left\{
\int\!\!d\bm{\omega}_n
d\bm{\omega}'_{n'}h(\bm{\omega}_n,\tilde{\bm{\omega}}_m)h^{\prime
*}(\bm{\omega}'_{n'},\tilde{\bm{\omega}}'_{m'})
\vphantom{\left(\prod_{j=1}^{J'}\right)}\right.\nonumber\\ 
& & \left.\langle
0_a|\left(\prod_{j=1}^{J'}a_j^{n'_j}(\bm{\omega}'_j)\right)
\left(\prod_{j=1}^Ja_j^{\dag n_j}(\bm{\omega}_j)\right)|0_a\rangle
\right\}\left(\prod_{k=1}^Kb_k^{\dag
m_k}(\tilde{\bm{\omega}}_k)\right)|0_b\rangle \langle
0_b|\left(\prod_{k=1}^{K'}
b_k^{m'_k}(\tilde{\bm{\omega}}'_k)\right)\nonumber \\ &= &
\delta_{JJ'}\left(\prod_{j=1}^Jn_j!\delta_{n_jn'_j}\right)\int\!\!
d\tilde{\bm{\omega}}_m d\tilde{\bm{\omega}}'_{m'}
\left\{ \int\!\!d\bm{\omega}_n\,
h(\bm{\omega}_n,\tilde{\bm{\omega}}_m)\left(
\prod_{j=1}^J\mathcal{S}(\bm{\omega}_j)\right) h^{\prime
*}(\bm{\omega}_{n},\tilde{\bm{\omega}}'_{m'}) \right\} \nonumber \\
& & \times\left(\prod_{k=1}^Kb_k^{\dag
m_k}(\tilde{\bm{\omega}}_k)\right)|0_b\rangle \langle
0_b|\left(\prod_{k=1}^{K'} b_k^{m'_k}(\tilde{\bm{\omega}}'_k)\right).
\end{eqnarray}
[The last equation follows from Lemma (B15) of Appendix~\ref{app:B}.] 
Because the $b$-creation operators commute among themselves, as do the
$b$-annihilation operators, there is one more symmetry:
\begin{eqnarray}\lefteqn{\mbox{Tr}_a(|\psi\rangle\langle\psi'|)}\nonumber
\\ &=&
\delta_{JJ'}\left(\prod_{j=1}^Jn_j!\delta_{n_jn'_j}\right)\int\!\!
d\tilde{\bm{\omega}}_m d\tilde{\bm{\omega}}'_{m'}\left\{\int\!\!d\bm{\omega}_n\left[\left(\prod_{k=1}^K\mathcal{S}
(\tilde{\bm{\omega}}_k)\right)\left(\prod_{j=1}^J\mathcal{S}
(\bm{\omega}_j)\right)h(\bm{\omega}_n,\tilde{\bm{\omega}}_m)\right]\right.
\nonumber \\ & &\times\left.\!\!
\left[\left(\prod_{k=1}^{K'}\mathcal{S}(\tilde{\bm{\omega}}'_k)\!\right)
\!\left(\prod_{j=1}^J\mathcal{S}(\bm{\omega}_j)\!\right) h^{\prime
*}(\bm{\omega}_{n},\tilde{\bm{\omega}}'_{m'}) \right]\right\}
\!\left(\prod_{k=1}^Kb_k^{\dag
m_k}(\tilde{\bm{\omega}}_k)\!\right)|0_b\rangle \langle
0_b|\left(\prod_{k=1}^{K'}
b_k^{m'_k}(\tilde{\bm{\omega}}'_k)\!\right)\!.\ \nonumber\\
\end{eqnarray}

\subsection{Bi-photons: excitation in each of two orthogonal modes}
For a state that exhibits a single photon in each of two orthogonal
modes, $a$ and $b$, whether in a single fiber or in different fibers,
the general form is
\begin{equation} (h\!:\!a^\dag b^\dag)|0\rangle, \label{eq:two}\end{equation}
where we extend our notation by defining
\begin{equation} 
 (h\!:\!a^\dag b^\dag) \defeq \int\!\! \int d\omega\,
d\tilde{\omega}\, h(\omega,\tilde{\omega})a^\dag(\omega)
b^\dag(\tilde{\omega}),
\label{eq:twoFac}
\end{equation} with the normalization requirement
that 
\begin{equation}
1 = \int\!\!\int d\omega\, d\tilde{\omega}\, |h(\omega,\tilde{\omega})|^2.
\end{equation} 
\noindent Note: the normalization requirement rules out
$h(\omega,\tilde{\omega})$ of the form
$f(\omega)\delta(\omega-\tilde{\omega})$.

Other states, including multi-mode, multi-photon states will
be introduced after we have the machinery of projections.

\section{Projections}\label{sec:3}
To deal efficiently with polarization-entangled states, we need to
characterize subspaces of states by the projections that leave them
invariant.  These are provided here.  It is instructive to compare and
contrast the projections for the function space of states used here
with the case of a single oscillator with its basis states
$|n\rangle$, $n = 0$, 1, \dots\ \cite{louisell}.  With our range of
frequencies there are countless 1-photon states (and countless $n$-photon
states), as described in Sec.\ \ref{subsec:2B}, in contrast
to the single $n$-photon state $|n\rangle$ for given $n$ of an
oscillator; however, there is a one-to-one correspondence between a
set of projection operators for photon number and the projections
$|n\rangle\langle n|$ for the oscillator.

For both the single oscillator and the states that are characterized
by functions of frequency, one writes
$|0\rangle \langle 0|$ for the vacuum projector.  To the oscillator
projection $|1\rangle\langle 1|$ corresponds the projection that leaves
invariant single-photon states while killing all states of more or fewer
photons: 
\begin{equation} P_1(a) = \int d\omega\, 
a^\dag(\omega) |0\rangle \langle 0| a(\omega).
\label{eq:P1}
\end{equation}  
It is easy to check that for any $f$ one gets $P_1(a)(a_f^\dag |0\rangle)
= a_f^\dag|0\rangle$, and that for any of the $n$-photon states
$|\psi_n\rangle$ with $n \ne 1$, one finds $P_1(a) |\psi_n\rangle = 0$.
Similarly the projection for two-photon states is
\begin{equation}
P_2(a) = \frac{1}{2}\int\!\!\int d\omega_1\, d\omega_2\,
a^\dag(\omega_1)a^\dag(\omega_2)|0\rangle\langle
0|a(\omega_1)a(\omega_2).
\label{eq:P2}
\end{equation} 
The general case is 
\begin{equation}
P_n(a) = \frac{1}{n!}\int\!\cdots\!\int d\omega_1 \cdots d\omega_n\,
a^\dag(\omega_1)\cdots a^\dag(\omega_n)|0\rangle\langle
0|a(\omega_1)\cdots a(\omega_n).
\label{eq:PnDef}\end{equation}
It follows that any state in the space of superpositions over
all $n$ is unchanged by the operator obtained by summing over
all $P_n(a)$, and hence this sum is the unit operator
\begin{equation}\sum_{n=0}^\infty P_n(a) = \mathbf{1}.
\label{eq:unit}
\end{equation}
Equation (\ref{eq:PnDef}) implies a relation among the projections for
adjacent values of $n$
\begin{equation}
P_{n+1}(a) = \frac{1}{n+1}\int d\omega\, a^\dag(\omega)P_n(a) a(\omega).
\label{eq:pn1}
\end{equation}
Although this mirrors the situation for a single oscillator, the
projections $P_n(a)$ {\em cannot} be expressed as the outer product of
a vector with its adjoint; indeed the $n$-photon subspaces are
infinite dimensional.

{}From the commutation Eqs.\ (\ref{eq:comm0}) and (\ref{eq:comm1}) it
follows that
\begin{equation}
P_n(a)\prod_{j=1}^m a^\dag(\omega_j)|0\rangle =
\delta_{n,m}\prod_{j=1}^m a^\dag(\omega_j)|0\rangle;
\label{eq:projN}
\end{equation}
furthermore, this holds if one inserts into both sides of the product
the same polynomial in operators for modes orthogonal to $a$.
Also from Eq.\ (\ref{eq:projN}) follows the relation
\begin{equation}\
P_n(a)(h\!:\!a^{\dag m})|0\rangle = \delta_{n,m}(h\!:\!a^{\dag m})|0
\rangle.
\label{eq:projNh}
\end{equation}

\subsection{Action of single-mode projections on multi-mode states}
In the context of two orthogonal modes $a$ and $b$, the symbol
$P_n(a)$ is re-used as shorthand for $P_n(a)\otimes \mathbf{1}_b$,
where $\mathbf{1}_b$ is the unit operator on the $b$-factor of a
tensor product. Two projections for differing values of $n$ for a
given mode are mutually orthogonal, but not when one projection is for
one mode and the other projection for another; \textit{e.g.}
$P_1(a_1)P_3(a_1)=0$ but $P_1(a_1)P_3(a_2) \ne 0$.  Similar remarks
apply to the case of more than two mutually orthogonal modes.

\subsection{Multi-mode $\protect\bm{n}$-photon projector}
To study polarization we will need the idea of an $n$-photon state
that can be distributed over two modes. We denote the projector for
$n$ photons distributed arbitrarily among modes $a_1$ and $a_2$ by
\begin{equation}
P_n(a_1,a_2) = \sum_{k=0}^n P_k(a_1)P_{n-k}(a_2).
\label{eq:Pdist}
\end{equation}

\noindent\textbf{Note}: All these projectors can be subdivided
into ``$+$'' and ``$-$'' parts, \textit{e.g.}\ 
\begin{equation} P_n(a) = P_{n,+}(a) + P_{n,-}(a).
\label{eq:split}
\end{equation}

\subsection{Number operator}
Similarly to the single oscillator case, one constructs an operator
that has as its expectation value for a given state the mean photon
number for that state.  This is the number operator for the mode
under discussion: 
\begin{equation}\hat{N}(a) \defeq  \sum_{n=0}^\infty
 n P_n(a).
\end{equation}  By use of Eqs.\
(\ref{eq:pn1}) and (\ref{eq:unit}), this is transformed into the more
convenient form:
\begin{eqnarray}
\hat{N}(a) & = & \sum_{n=1}^\infty n P_n(a) =
\sum_{n=1}^\infty n\, \frac{1}{n}\int d\omega\,
a^\dag(\omega)P_{n-1}(a)\, a(\omega) \nonumber \\ & = & \int d\omega\,
a^\dag(\omega)\left(\sum_{n=1}^\infty P_{n-1}(a)\right) a(\omega)
= \int d\omega\, a^\dag(\omega) a(\omega).
\label{eq:number}
\end{eqnarray}
It is also useful to express the number operator for the ``$+$''
and ``$-$'' directions:
\begin{equation}
\hat{N}_\pm(a) = \int_0^\infty d\omega\, a_\pm^\dag(\omega)
a_\pm(\omega).
\label{eq:Ndir}
\end{equation}

For a space (or subspace) spanned by two modes $a_1$ and $a_2$, define
\begin{equation}
\hat{N}(a_1,a_2) = \sum_{n=0}^\infty n P_n(a_1,a_2).
\end{equation} 
As sketched in Lemma (\ref{eq:b7}) of Appendix~\ref{app:B}, a calculation
similar to that for one mode shows
\begin{equation}
\hat{N}(a_1,a_2) = \int d\omega\, [a_1^\dag(\omega)
a_1(\omega)+ a_2^\dag(\omega) a_2(\omega)].
\label{eq:N2def}
\end{equation}  
Here again, we can pick out directions, as in Eq.\ (\ref{eq:Ndir}).
There is no difficulty in extending to more than two modes.

\section{Loss and frequency dispersion}\label{sec:4}

We partition the modes to be analyzed into ``desired modes'' and
``extraneous modes,'' as illustrated in Fig.\ \ref{fig:2}.  By
considering coupling of desired modes to extraneous loss modes, one
can model a wide variety of loss mechanisms.  The choice of model
for loss is tied to the choice of model of detection.

To model loss of a mode to be detected by a binary detector,
neglecting memory effects in the detector, for now we succumb to the
charm of simplicity in the approach of Mandel [\onlinecite{mandel},
p.~640].  In this approach a mode $b'$ prior to loss is related to a mode
$b$ after loss by the following equation for the respective annihilation
operators:
\begin{equation} b'(\omega) = \eta_{\rm loss}(\omega) b(\omega) +
  (1-|\eta_{\rm loss}(\omega)|^2)^{1/2}  c(\omega),
\label{eq:loss}
\end{equation}
where $c$ expresses an undetected mode into which all the $b'$-energy
spills except for a fraction.  In the case where $\eta_{\rm loss}$ is
independent of frequency, this un-lost fraction is just $|\eta_{\rm
loss}|^2$.

An alternative is to consider the coupling of desired modes to a heat
bath. This leads to so-called master equations.  The simplest form
that expresses the essential features makes the Markhoff approximation
[\onlinecite{louisell}, p.~347, Eq.~(6.2.61)].  With further
simplifications, including that of a zero-temperature heat bath, we
obtain the time behavior of a density operator that at time $t_0$
expresses the state
$a_f^\dag|0\rangle$ to be, in the Schr\"odinger picture (SP), a density
operator of the form\looseness=-1
\begin{eqnarray}\rho_f & = & \int\!\!\int d\omega\, d\omega'\, 
\exp\{-[\gamma(\omega)+\gamma(\omega')] t\}
\,f(\omega)f^*(\omega')e^{i(\omega - \omega')t}a^\dag(\omega)|0\rangle
\langle 0|a(\omega') \nonumber \\ & &{} + \left(1-\int d\omega
\,\exp\{-[\gamma(\omega)+\gamma(\omega')]
t\}\, |f(\omega)|^2\right)|0\rangle \langle 0|.
\label{eq:diss}
\end{eqnarray}

\noindent [Need to work this out for multi-photon states.]

\subsection{Loss cannot evade ``no cloning''}

I believe (and need to check) that coupling of desired modes to
thermal modes cannot make two states more distinguishable.  More
formally, suppose the desired modes
are $A$ and the extraneous modes are $B$, and the in-state is
expressed by either some tensor product $\rho_{A,{\rm in},1} \otimes
\rho_{B,{\rm in}}$ or by $\rho_{A,{\rm in},2} \otimes \rho_{B,{\rm
in}}$.  After a unitary evolution $U$ acting on the tensor-product
space of $A$ and $B$, the out-state is $U\rho_{A,{\rm in},j} \otimes
\rho_B U^\dag$, with $j = 1$ or~2.  The reduced density matrix for $A$
is obtained by the partial trace over $B$:
\begin{equation} \rho_{A,{\rm out},j} = \mbox{Tr}_B(U\rho_{A,{\rm
      in},j} \otimes \rho_B U^\dag).
\end{equation}
(Note that, unlike traces, partial traces of a product depend on the
order of the factors.)  My guess is that in all cases
\begin{equation}
\mbox{Tr}(\rho_{A,{\rm out},1}^{1/2}\rho_{A,{\rm out},2}^{1/2}) \ge
\mbox{Tr}(\rho_{A,{\rm in},1}^{1/2}\rho_{A,{\rm in},2}^{1/2}).
\end{equation}

{}For this it is necessary and sufficient to prove for all density
operators $\rho_1$ and $\rho_2$ acting on the tensor-product space of
$A$ and $B$ that
\begin{equation}
\mbox{Tr}_A[(\mbox{Tr}_B\,\rho_1)^{1/2}(\mbox{Tr}_B\,\rho_2)^{1/2}] \ge
\mbox{Tr}(\rho_1^{1/2}\rho_2^{1/2}).
\end{equation}

\section{Local quantum fields}\label{sec:5}

Improper operators introduced so far have been integrated over
frequency to produce proper operators.  We want also to construct
proper operators by integrating over space and/or time, which leads us
to take Fourier transforms, as outlined in Appendix~\ref{app:D}.  In
analogy with quantum electrodynamics for propagation in vacuum, we
introduce (improper) local annihilation field operators in the
Heisenberg picture: 
\begin{eqnarray}a_+(x,t) & = & \frac{1}{\sqrt{2\pi}}\int_0^\infty
d\omega\, a_+(\omega)e^{-i[\omega t-k(\omega)x]}, \nonumber \\
a_-(x,t) & = & \frac{1}{\sqrt{2\pi}}\int_0^\infty
d\omega\, a_-(\omega)e^{-i[\omega t+ k(\omega)x]}, 
\label{eq:loca}
\end{eqnarray}
where $k(\omega)$ is an experimentally determined propagation
factor, defined so that
$\mbox{sgn}\, k(\omega) = \mbox{sgn}\, \omega$.
We assume 
\begin{equation} 
\hbox to .9\textwidth{\hbox to .15\parindent{}1.\hfil $k(-\omega) =
-k(\omega).$\hfil}
\end{equation}
\begin{equation}
\hbox to .9\textwidth{\hbox to .15\parindent{}2.\hfil $\lim_{\omega
\rightarrow
\infty} {\displaystyle \frac{k(\omega)}{\omega}} > 0.$\hfil}
\end{equation}
\begin{equation}
\hbox to .9\textwidth{\hbox to .15\parindent{}3.\hfil
${\displaystyle \frac{dk(\omega)}{d\omega} >
0.}$\hfil}
\end{equation}

\vskip8pt
4. Kramers-Kronig relations connect this $k(\omega)$ to the loss
coefficient $\gamma(\omega)$ of Eq.\ (\ref{eq:diss})
\cite{jackson}.\looseness=1

The operator $a_\pm^\dag(x,t)$ is `improper' in that it takes a
normalized state to an unnormalizable state; a proper operator can be
defined by averaging $a_\pm^\dag(x,t)$ over a spacetime region.
(This averaging takes place automatically in time-dependent
perturbation theory [\onlinecite{louisell}, p.~257].)\ \
Drawing on Eq.\ (\ref{eq:adef}), we can define
\begin{eqnarray}a(x,t) & \defeq  & a_+(x,t) + a_-(x,t)
=\frac{1}{\sqrt{2\pi}}\int_{-\infty}^\infty
d\omega\, a(\omega) e^{-i [|\omega| t- k(\omega)x]},
\end{eqnarray}
and from Eqs.\ (\ref{eq:loca}) and (\ref{eq:comm0}) follows the
commutation relation
\begin{equation}
[a(x,t),a^\dag(\omega)] = \frac{1}{\sqrt{2\pi}}\exp\{-i[|\omega| t
-k(\omega)x]\}.
\end{equation}

\subsection{Temporally local hermitian fields}
{}From the non-hermitian operator $a(x,t)$ can be constructed the two
non-commuting hermitian quadrature operators
\begin{eqnarray} q(x,t) & \defeq  &
a^\dag(x,t)+a(x,t) \nonumber \\ & = &
\frac{1}{\sqrt{2\pi}}\int_{-\infty}^\infty d\omega\, \{a^\dag(\omega)
e^{i [|\omega| t- k(\omega)x]} + a(\omega)
e^{-i [|\omega| t- k(\omega)x]}\}, \\ p(x,t) & \defeq  & i[a^\dag(x,t) - a(x,t)] \nonumber \\ & = &
\frac{i}{\sqrt{2\pi}}\int_{-\infty}^\infty d\omega\, 
\{a^\dag(\omega)
e^{i [|\omega| t- k(\omega)x]} - a(\omega)
e^{-i [|\omega| t- k(\omega)x]}\}. 
\end{eqnarray}
Note $p(x,t)$ is like Louisell's voltage operator.
Fourier-transforms show
\begin{equation}\int_{-\infty}^\infty dt\, a_\pm^\dag(x,t)a_\pm(x,t)
= \int_0^\infty d\omega\, a_\pm^\dag(\omega)a_\pm(\omega).
\label{eq:dt}
\end{equation} This will be used in constructing operators to model
detection.

\subsection{Time, space, and dispersion}
Because the variation of $k(\omega)$ with $\omega$ is non-linear, the
operator that is convenient for space localization must differ
from that which is convenient for
time localization.  Consider operators such as the hamiltonian and the
number operator (shortly to be introduced) of the form $\hat{O}= \int
d\omega\, f(\omega) a^\dag(\omega)a(\omega)$.  In the case of the
number operator defined as $\hat{N}_\pm(a)= \int_0^\infty d\omega\,
a_\pm^\dag(\omega)a_\pm(\omega) = \int dt\,
a_\pm^\dag(x,t)a_\pm(x,t)$, the local operator
$a_\pm^\dag(x,t)a_\pm(x,t)$ acts as a kind of density in time for
$\hat{N}_\pm(a)$; however, the space integral of
$a_\pm^\dag(x,t)a_\pm(x,t)$ is something else:\looseness=-1
\begin{equation}
\int^\infty_{-\infty} dx\, a_\pm^\dag(x,t)a_\pm(x,t) =
\int_0^\infty d\omega_1\!\int_0^\infty d\omega_2 \,
a_\pm^\dag(\omega_1)a_\pm(\omega_2)
e^{i(\omega_1-\omega_2)t}\delta[k(\omega_1)-k(\omega_2)],
\end{equation}
which, with the relation
\begin{equation}\delta[k(\omega)-k(\omega')] =
    \left|\frac{dk(\omega)}{d\omega}\right|^{-1}
\delta(\omega - \omega'),
\end{equation} becomes
\begin{equation} \int^\infty_{-\infty} dx\, a_\pm^\dag(x,t)a_\pm(x,t)  =
  \int_0^\infty d\omega\,
  a_\pm^\dag(\omega)a_\pm(\omega)\left|\frac{dk(\omega)}{d\omega}
\right|^{-1}.
\end{equation} 
Note that the ``extra factor'' $1/|dk/d\omega|$ expresses
a group velocity \cite{brillouin}.  

With this in mind, we construct an energy density operator
$a'_\pm(x,t)$ by
\begin{equation}
a'_\pm(x,t) \defeq  \frac{1}{\sqrt{2\pi}}\int_0^\infty
d\omega \left(\omega\,
\frac{dk(\omega)}{d\omega}\right)^{1/2}a_\pm(\omega)e^{-i[\omega
	t \mp k(\omega)x]}.
\end{equation}
One can check that this has the property
\begin{equation}
\int_{-\infty}^\infty dx\, a'^\dag_\pm(x,t)a'_\pm(x,t) = \int_0^\infty
d\omega\,|\omega| a_\pm^\dag(\omega)a_\pm(\omega) = H/\hbar.
\label{eq:H1}
\end{equation}
There are lots of other possibilities.  For instance, define a
field 
\begin{equation}
\tilde{a}(x,t) \defeq 
\frac{1}{\sqrt{2\pi}}\int_{-\infty}^\infty d\omega\ 
\frac{dk(\omega)}{d\omega}\, a(\omega) e^{-i [|\omega| t-
k(\omega)x]};
\end{equation}
then \begin{equation}
H/\hbar = i\int_{-\infty}^\infty dx\, \tilde{a}^\dag(x,t)\,\frac{\partial
  a(x,t)}{\partial t}, 
\label{eq:H2}\end{equation} 
so that the two distinct fields $a$ and $\tilde{a}$ both
enter.  

In order to deal with finite fibers, we explore the operator obtained
by making the limits of integration over $x$ finite; however, the
resulting integral is no longer independent of $t$.  Further, while the
integral obtained from putting finite $x$-limits in (\ref{eq:H1}) is
at least hermitian, even this modest property fails for the integral
obtained by putting finite $x$-limits in (\ref{eq:H2}).  Nonetheless,
in some cases such a truncated operator can be a useful approximation
to the hamiltonian.

\subsection{Projections in terms of local operators}
It will be interesting to examine approximations to projections in
terms of operators that are, so to speak, confined in time and space.
For reference, here we express some projections in terms of local
operators.  From Eq.\ (\ref{eq:P1}) and the inverse
Fourier transform in time of Eq.\ (\ref{eq:loca}), one computes
\begin{equation}
P_1(a) = \int^\infty_{-\infty} dt\,a^\dag(x,t)|0\rangle \langle
0|a(x,t)
\end{equation}
(independent of $x$).
Similarly, from Eq.\ (\ref{eq:P2}) one computes
\begin{equation}
P_2(a) =
\int^\infty_{-\infty} dt\int^\infty_{-\infty} dt'\,a^\dag(x,t)
a^\dag(x',t')|0\rangle \langle 0|a(x,t)a(x',t'),
\end{equation}
and one can keep on going.  This procedure also works
to express the number operator of Eq.\ (\ref{eq:number}) in terms
of local operators:
\begin{eqnarray}
\hat{N}(a) & = &  \int d\omega\, a^\dag(\omega) a(\omega)
=\int^\infty_{-\infty} dt\, a^\dag(x,t)a(x,t).
\label{eq:numberLoc}
\end{eqnarray}

\section{Scattering matrix}\label{sec:6}
Couplers and other networks can be analyzed in terms of a scattering
matrix.  For this, one supposes that there is time $t_1$ before which
the network acts as a set of uncoupled fibers and perhaps a quantum
memory uncoupled to these fibers; one supposes a later time $t_2$
after which the interaction is over, so that again one has a set of
uncoupled fibers and a quantum memory in isolation.  Prior to $t_1$,
the light is modeled by an in-state as an integral over frequency of a
polynomial in in-mode creation operators acting on the vacuum.  For
time $t > t_2$ the light is modeled by an out-state, again as an
integral over frequency of a polynomial in out-mode creation operators
acting on the vacuum.

For example, to analyze a coupler, we typically express state
preparation in terms of in-state creation operators while we
express detection in terms of out-state creation operators.  Hence
calculating the probability of an outcome calls for expressing
in-state operators in terms of out-state operators, or vice-versa,
according to whichever is more convenient.

The use of scattering theory in this context is to relate the
creation operators for in-state modes to creation operators for
out-state modes, and this relation is constrained to be a unitary
transformation.  Because of unitarity, commutation relations are
preserved and so are inner products.  A complication is that unitarity
holds only for a transform over all the modes involved, ``extraneous''
as well as ``desired.''

In general the creation operator for an out-mode at one frequency
depends on creation operators for in-modes at all frequencies;
i.e., the operator mixes frequencies, as in parametric down conversion.
For linear networks we get the major simplification that out-mode
operators for a given frequency $\omega$ are unitary transforms of
in-mode operators for the same frequency.

After a scattering transformation is obtained, either by guessing
or by calculating it from some more detailed model, one can
check whether the in-states and out-states are consistent with
particular values of $t_1$ and $t_2$.  (For a network with strong
internal reflections, the interval $t_2 - t_1$ must be large enough
to allow reverberations to die out.)

\subsection{Network without frequency mixing}
For the special case of a network that is linear in that it mixes no
frequencies, if the extraneous modes are accounted for, then the
out-state annihilation operators $a_{j,{\rm out}}(\omega)$, $j = 1$, 2,
\dots, are some unitary transform of the in-state annihilation
operators.  That is, let
\begin{equation}
\vec{a}_{\rm out} = \left[\begin{array}{c}a_{1,{\rm out}} \\
a_{2,{\rm out}} \\
\vdots \\
\end{array}\right],
\end{equation} 
and similarly define $\vec{a}_{\rm in}$.  Then there is some
frequency-dependent, unitary matrix $U(\omega)$, having dimension
equal to the number of out-modes (by assumption equal to the number of
in-modes), such that
\begin{equation}
\vec{a}_{\rm out}(\omega) = U(\omega) \vec{a}_{\rm in}(\omega).
\label{eq:scatt}
\end{equation}
Taking hermitian conjugates and multiplying the result by $U(\omega)$
on the right gives a relation between the row vectors  $\vec{a}^\dag_{\rm
  out}$
and  $\vec{a}^\dag_{\rm in}$:
\begin{equation}\vec{a}^\dag_{\rm out} = \vec{a}^\dag_{\rm
    in}U^\dag(\omega).
\label{eq:scattdag}
\end{equation}

To understand the application of scattering formalism to modeling
fiber interactions illustrated in Fig.\ \ref{fig:1}, first model light
pulses as quantum states expressed by creation operators acting on
vacuum.  We think of an interaction that, to some approximation, starts
at $t_1$ or later and is complete by $t_2$ or earlier. Prior to $t_1$, the
incoming light is modeled as involving only in-state creation
operators.  After $t_2$, the light is modeled as a different state,
involving only out-state creation operators, related to the in-state
creation operators by an equation of the form (\ref{eq:scattdag}).

\section{Polarized and entangled light states}\label{sec:7}

It is possible to make a fiber, for instance with an elliptical cross
section, that propagates only a single mode; however, the fibers used
in the Quantum Network all carry two polarization modes.  We follow
convention by misnaming fibers, so that when we say a `single-mode'
fiber we mean a fiber that propagates not a single spatial mode $a$ but
one that propagates two polarized modes $a_1$ and $a_2$, mutually
orthogonal in that the $[a_1,a_2^\dag] = 0$, with propagation
constants $k_{a1}(\omega)$ and $k_{a2}(\omega)$, respectively. This
means that for a network (as in Fig.\ \ref{fig:1}) with $n$ `single-mode'
fibers there are $2n$ in-modes and $2n$ out-modes, so that, neglecting
extraneous modes, the single-frequency scattering matrix $U(\omega)$
has dimension $2n$.

(For special materials, one may need to separate out the two
directions to allow $k_{aj+} \neq k_{aj-}$.)  Note that a linear
superposition of the two modes with different propagation constants
has no definable propagation constant.

In describing modeling approaches for fibers that support two
polarizations we make (and try to state) various simplifying
assumptions; throughout we assume frequency conservation.  Taking
polarization into account doubles the number of modes relative to the
number of so-called single-mode fibers, for the simple reason that
these are misnamed.

For a polarized fiber having orthogonal modes $a_1$ and $a_2$, the
form of a single-photon state is [see Eq.\ (\ref{eq:mulmode})]:
\begin{equation} |\mbox{1-photon}\rangle = (c_1 a_{1,f}
+ c_2 a_{2,g})^\dag|0\rangle, 
\end{equation}
for any normalized $f$ and $g$, assuming numerical constants $|c_1|^2
+ |c_2|^2 = 1$.  The same operator generates the coherent state
of the form of Eq.\ (\ref{eq:coh}):
\begin{eqnarray}|\alpha,(c_1 a_{1,f}
+ c_2 a_{2,g})\rangle & \defeq  &
e^{-|\alpha|^2/2}e^{\alpha(c_1 a_{1,f} + c_2 a_{2,g})^\dag}|0\rangle
\nonumber
  \\ & = & |c_1 \alpha,a_{1,f}\rangle \otimes |c_2 \alpha,a_{2,g}\rangle,
\end{eqnarray}
where the second equation follows from $\exp[(c_1 a_{1,f}
+ c_2 a_{2,g})^\dag] = \exp(c_1 a_{1,f}^\dag)\exp(c_2 a_{2,g})^\dag$,
which is a consequence of the commutativity $[a_1^\dag,a_2^\dag] = 0$.

[**The whole business seems to assume a reference position $x = 0$;
other values of $x$ involve $x$-dependent phase factors.]

\subsection{Fiber splice (without extraneous modes)}

\begin{figure}[t] 
\begin{center}
\includegraphics[width=4.75in]{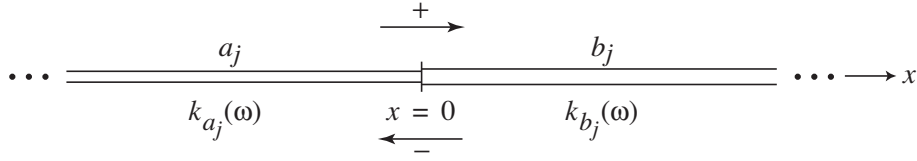}
\end{center}
\vspace{-15pt}\caption{Splice in fibers of different type.}\label{fig:3}  
\end{figure}

Consider the situation shown in Fig.\ \ref{fig:3}, where a fiber $a$
on the left is spliced to a fiber $b$ on the right.  We suppose the
splice is centered at $x=0$. For such a heterogeneous spliced fiber,
the local field operator $a(x,t)$ cannot satisfy a field equation that
exhibits translational symmetry; instead, the field equation stems
from a hamiltonian that has a change at the splice; the situation is
reminiscent of a potential problem in quantum mechanics.

Without trying to model the details, and neglecting coupling to
extraneous modes, the effect of the splice is to convert in-mode
operators to out-mode operators, as expressed by a unitary matrix
$U(\omega)$.  Recognizing polarizations, we have modes $a_1$ and $a_2$
in the $a$-fiber and modes $b_1$ and $b_2$ in the $b$-fiber, with
propagation constants $k_{a_1}$, $k_{a_2}$, $k_{b_1}$, and $k_{b_2}$,
respectively.  This makes the four in-modes $a_{j+}$ and $b_{j-}$, $j
= 1,2$, along with the four out-modes $a_{j-}$ and $b_{j+}$; hence if
we neglect loss, the scattering matrix $U(\omega)$ is 4-by-4:
\begin{equation}
[a_{1-}^\dag(\omega),\,b_{1+}^\dag(\omega),\,a_{2-}^\dag(\omega),\,
b_{2+}^\dag(\omega)] = [b_{1-}^\dag(\omega),\,a_{1+}^\dag(\omega),\,
b_{2-}^\dag(\omega),\,a_{2+}^\dag(\omega)]U^\dag(\omega).
\label{eq:splice}
\end{equation}

Under the simplifying assumption that the splice has no coupling
between polarizations, the equation factors into two 2-by-2 pieces,
one piece for each polarization.  With $U(\omega)$ defined this way, the
perfectly homogeneous situation in which $k_{a_j}(\omega) =
k_{b_j}(\omega)$, thus making the splice invisible, corresponds to
$U(\omega) = 1$.  

(To account for loss, we can add extraneous dimensions and then trace
them out, thereby getting a matrix that is not unitary, so that the
out-power can be less than the in-power.) [*For single-frequency
matrices, we use power rather than energy, because energy is definable
only for non-zero bandwidth.]

\subsection{Coupler}
\begin{figure}[t] 
\begin{center}
\includegraphics[width=2.75in]{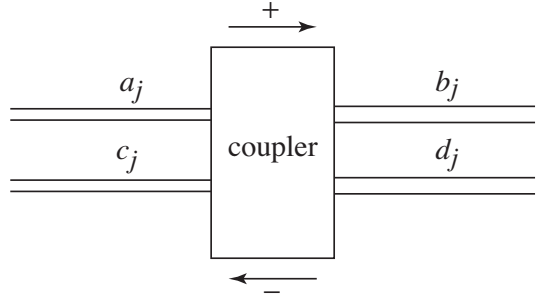}
\end{center}
\vspace{-25pt}
\caption{Fiber coupler.}\label{fig:4}
\end{figure}

Fiber couplers, analogous to beam splitters, have four fibers,
$a_j,b_j,c_j,d_j$, with $j = 1, 2$ (see Fig.\ \ref{fig:4}).  Each of the
eight modes comes in two directions, ``$+$'' and ``$-$''.
Without loss, $U(\omega) \in \text{SU}(8)$.

\subsection{Entangled states}
Whether entangled in frequency or in polarization or in both,
entangled states  have to do with tensor products of vector spaces.
Let a vector space $V$ be a tensor product of vector spaces
$V_\alpha$, where $\alpha$ ranges over some index set.  Relative to this
factorization, a generic vector in
$v
\in V$ is a sum of tensor products of vectors $v_\alpha \in V_\alpha$;
unless it can be written as a single product, not a sum of products,
$v$ is called \textit{entangled} (relative to the factorization).

We stress \textit{relative to a factorization} because vector spaces for
light states can be factored in more than one way, and to speak
sensibly of `entangled states' one must know which factorization is
meant.  The vector space for a single mode involves tensor products
over subspaces for frequency bands; this factorization is relevant to
frequency-entangled states.  Within any frequency band, there is an
infinite tensor product over single frequencies, which will not be
used here in speaking of `entanglement'.  Polarization-entanglement
involves a vector space for several modes, factored by mode.

Frequency-preserving, unitary transformations of the single-frequency
creation operators carry an in-state that is a tensor product of
broad-band coherent states to an out-state that is also a tensor
product of broad-band coherent states.  This is a special
property; indeed the only case I know in which it works is that of
coherent states. [* Work out proof] (In generic cases, a tensor product of
single-photon in-states is carried to an entangled, multi-photon
state.)  This makes broad-band coherent states of special interest
as an (overcomplete) basis for studying light in the context
of frequency-preserving splices and couplers.  See Appendix \ref{app:E}.

\subsection{Polarization-entangled states}
A polarization-entangled state involves four modes; we think of two
orthogonal modes $a_1$ and $a_2$ in a left fiber to Alice and two
orthogonal modes $b_1$ and $b_2$ in a right fiber to Bob.  (One can
think of 1 as horizontal and 2 as vertical.)  Then there are four
creation operators $a_1^\dag(\omega)$, $a_2^\dag(\omega)$,
$b_1^\dag(\omega)$, and $b_2^\dag(\omega)$; everything commutes except
that $[a_i(\omega),a_j^\dag(\omega')] = \delta_{i,j}\delta(\omega
-\omega')$, where $\delta_{i,j}= 1$ if $j=i$ and otherwise is 0.

The defining property of a state $|\phi\rangle$ characterized by a
single $a$-photon combined with a single $b$-photon is
\begin{equation}
P_1(a_1,a_2)P_1(b_1,b_2)|\phi\rangle = |\phi\rangle. 
\end{equation}
A necessary and sufficient condition for this is that $|\phi\rangle$
have the form
\begin{equation}
|\phi\rangle = \sum_{i,j = 1}^2 C_{i,j} (h\!:\!a_i^\dag b_j^\dag)|
 0\rangle,
\label{eq:biPhot}
\end{equation}  where
\begin{equation}
\sum_{i,j = 1}^2 |C_{i,j}|^2 = 1,
\end{equation}
where we use the definition stated in Eq.\ (\ref{eq:twoFac}),
and the four otherwise arbitrary functions $h_{i,j}$ satisfy the
normalization condition (in which no symmetry is assumed): 
\begin{equation}
\int\!\!\int d\omega\, d\tilde{\omega}\,
|h_{i,j}(\omega,\tilde{\omega})|^2 = 1.
\label{eq:norm2}
\end{equation}

To think about quantum key distribution, we will need to consider
states of both fewer and more photons.  A state with $m$ photons in
$a$-modes, $j$ of which are in mode $a_1$, and with $n$ photons in
$b$-modes, $k$ of which are in mode $b_1$, has the form
\begin{eqnarray} \lefteqn{\frac{1}{\sqrt{j!(m-j)!\,k!(n-k)!}}\,
(a_1^ja_2^{m-j}b_1^kb_2^{n-k})_h^\dag|0\rangle }\quad\nonumber \\
&=& \frac{1}{\sqrt{j!(m-j)!\,k!(n-k)!}}  \int\!\cdots\!\int
d\omega_1 d\tilde{\omega}_1\cdots d\omega_n
d\tilde{\omega}_n\,h(\omega_1,\tilde{\omega}_1,\ldots,\omega_n,\tilde{\omega}_n)
\nonumber \\ & &{}\times a_1^\dag(\omega_1)\cdots
a_1^\dag(\omega_j)a_2^\dag(\omega_{j+1})\cdots a_2^\dag(\omega_m)
 b_1^\dag(\tilde{\omega}_1)\cdots
b_1^\dag(\tilde{\omega}_k) b_2^\dag(\tilde{\omega}_{k+1})\cdots
b_2^\dag(\tilde{\omega}_n)|0\rangle.\qquad
\label{eq:multst}
\end{eqnarray}  
There is no loss of generality in requiring $h$ to be symmetric under
each interchange $\omega_j \leftrightarrow \omega_k$ for which
$\omega_j$ and $\omega_k$ pertain to the same mode. When $h$ with this
symmetry is normalized as in Eq.\ (\ref{eq:nnorm}), the state defined
by Eq.\ (\ref{eq:multst}) has unit norm. The most general state is a
weighted sum of such states.

\section{Detection}\label{sec:8}

A light detector as used in an experimental setup will be modeled by a
positive operator-valued measure (POVM), $\{M_j\}$, so the probability of
outcome $j$ is $\text{Tr}(M_j\rho)$, where $\rho$ is a density operator,
and $\sum_j M_j = 1$ with $M_j \ge 0$.  In the simplest models to be
discussed here, the density operator is for the light; in more complex
models discussed elsewhere \cite{0404113}, the density operator can be
for a pure or mixed state of not only light to be detected but also probe
particles.  One could also include the generation of backward-propagating
light, but here we do not get into this level of complication.

\subsection{Simple examples}

\noindent Example 2: As a first theoretical example, a detector as
narrow-band as possible that will register outcome 1 with certainty,
given the one-photon state $a^\dag_g|0\rangle$, has $M_1 = a^\dag_g
|0\rangle \langle 0|a_g$.  This `filtered' detector discriminates well
against any other one-photon state $a^\dag_f|0\rangle$ if $f$ is unlike
$g$; i.e., the probability for outcome 1 for a state
$a^\dag_f|0\rangle$ is $\mbox{Tr}(a^\dag_g |0\rangle \langle 0|a_g
a^\dag_f|0\rangle \langle 0|a_f) = |\langle 0|a_g a^\dag_f|0\rangle|^2
= |\int d\omega\, g^*(\omega)f(\omega)|^2$.

Given a set of mutually orthogonal functions $g_j$, $j = 1$, \dots, $N$,
there can be a detector with $N$ possible outcomes, with $M_j =
a^\dag_{g_j} |0\rangle \langle 0|a_{g_j}$.  Thus a single-photon
detector is certainly not restricted to giving a yes-no outcome, and
discrimination among different single-photon states is possible.
Herein lies a caution to designers who hope that distinct lasers
generate distinct states that differ in polarization. 
\vspace{3pt}

\noindent Example 3: An example of a model of detectors preceded by
filters used for joint detection of a two-mode state is $M_1 =
a^\dag_f b^\dag_g |0\rangle \langle 0|b_g a_f$.  Then outcome 1
would be described as `finding the state $a^\dag_f |0\rangle$ in
measuring mode $a$ jointly with finding the state $b^\dag_g |0\rangle$
in measuring mode~$b$.'  For the two-mode state (\ref{eq:two}), the
probability of this outcome is readily calculated to be $|\int\!\!\int
d\omega\, d\omega'\, f^*(\omega)g^*(\omega')h(\omega,\omega')|^2$.

\vspace{3pt}

\noindent Example 4: An interesting and perhaps novel application of
probe particles involves detection at two coordinated locations using
probe particles that have previously become entangled.  This can
produce an example that swaps the detection operator $M_1$ and the
state of Example 2, leaving the probability invariant.  I.e., the same
probabilities and outcomes arise from measuring an entangled
state with an unentangled detector as from measuring an unentangled state
with an entangled detector \cite{0404113}.

\subsection{Model of APD detector for quantum cryptography}

For multi-photon states, a simple approach is to model a detector as a
POVM for a state involving only the light to be detected.  This
assumes that memory effects in the detector, such as those often
attributed to trapped carriers in photo-diodes, are insignificant (for
instance because the network design enforces enough hardware dead
time); it also assumes teetering in the detector plays no significant
role \cite{0404113}.  For simplicity, here we add the assumption that
the detector and its pulse-shaping circuitry choose an outcome of
`yes' or `no' without any additional detail.  With these assumptions,
all one has left is dark-count and efficiency; one obtains a class of
detector models that respond to $n$ photon states according to
\begin{equation}
\Pr(\mbox{Detect}|n\mbox{-photon state})  =  d_n,
\end{equation}
with $0 \le d_n \le 1$.  With the additional assumption that the
detector is flat in its frequency response over the bandwidth of the
light pulses to be detected, the detection operator that generates
these probabilities is
\begin{equation}
M_1(a) = \sum_{n=0}^\infty \,d_n P_{n,+}(a),
\end{equation}
where $P_n(a)$ is defined in Eq.\ (\ref{eq:PnDef}) and we use
$a_+(\omega)$ in place of $a(\omega)$, so the integrals are over
positive frequencies only.  [For a detector of both polarization modes
$a_1$ and $a_2$ of a fiber, one has instead,
\begin{equation}
M_1(a_1,a_2) = \sum_{n=0}^\infty \, d_n P_{n,+}(a_1,a_2),
\end{equation} with $P_{n,+}(a_1,a_2)$ defined by Eq.\ (\ref{eq:Pdist})
with $a_+(\omega)$ in place of $a(\omega)$.]

To model the use of  avalanche photo-diode (APD) detectors in the
DARPA Quantum Network, we explore in detail a specialization of
this type of model.  We will use this model in Sec.\ \ref{sec:9} to
study the variation in detection statistics as the energy of
transmitted pulses is raised above that of a single-photon state.

An APD detector for an optical fiber responds to both polarization
modes of the fiber; however, by placing a polarizing beam splitter
before the detector, one can effectively eliminate the light from one
mode.  We assume this case, so that only one polarization mode is
relevant.  Then we suppose that the probability of the detector
responding to any single-photon state is $\eta_{\rm det}$, assumed
independent of frequencies over the range of frequencies relevant to
the light state.  We need to model the probability of detecting
multi-photon states.  By the assumption already made, the probability
of failing to detect any single-photon state is $1- \eta_{\rm det}$.
Assume that the probability of failing to respond to an $n$-photon
state is just the single-photon failure probability raised to the
$n$-th power: $(1-\eta_{\rm det})^n$.  Then the probability assigned
by this model to registering a detection, given an $n$-photon state
and temporarily ignoring dark counts, is
\begin{equation}
\Pr(\mbox{Detect}|n\mbox{-photon state})  =  1 - (1-\eta_{\rm
  det})^n .\label{eq:exact} 
 \end{equation}
To allow for dark counts we replace this by
\begin{equation}
\Pr(\mbox{Detect}|n\mbox{-photon state}) = 1 - (1-p_{\rm
  dark})(1-\eta_{\rm det})^n .\label{eq:exactDk}
 \end{equation}
To deal with states that are superpositions over varying numbers of
photons, we recall that the operator for an ideal detector that
responds always to any $n$-photon state, with no probability of
detection for a state that has no $n$-photon component, is the
projection $P_{n,\pm}$ defined in Sec.~\ref{sec:5}.  The following
detection operator $M_1(a)$ invokes $P_{n,\pm}$ to provide the desired
probabilities for a detector of a single $a$-mode \cite{perina}:
\begin{equation}
M_1(a) = \sum_{n=0}^\infty \,[1 - (1-p_{\rm dark}(a))(1-\eta_{\rm
  det})^n]P_{n,+}(a),
\label{eq:M1}
\end{equation}
where $P_n(a)$ is defined in Eq.\ (\ref{eq:PnDef}) and we use
$a_+(\omega)$ in place of $a(\omega)$, so the integrals are over
positive frequencies only.  [For a detector of both polarization modes
$a_1$ and $a_2$ of a fiber, one has instead,
\begin{equation}
M_1(a_1,a_2) = \sum_{n=0}^\infty \,[1 - (1-p_{\rm
dark}(a_1,a_2))(1-\eta_{\rm det})^n]P_{n,+}(a_1,a_2),
\label{eq:m1old}
\end{equation} with $P_{n,+}(a_1,a_2)$ defined by Eq.\ (\ref{eq:Pdist})
with $a_+(\omega)$ in place of $a(\omega)$.]

As applied to states with photon-number components negligible except
when $\eta_{\rm det}n \ll 1$, we notice
\begin{equation}1 - (1 - p_{\rm dark})(1-\eta_{\rm det})^n \approx
p_{\rm dark} + (1 - p_{\rm dark}) \eta_{\rm det} n,
\end{equation} from which we obtain for this case an approximation
that simplifies Eq.\ (\ref{eq:M1})
\begin{eqnarray}
M_1(a) & \approx & \sum_{n=0}^\infty \,[p_{\rm dark}(a) + (1-p_{\rm
  dark}(a))\eta_{\rm det} n]P_{n,+}(a) \nonumber \\
& = & p_{\rm dark}(a) + (1-p_{\rm dark}(a))\eta_{\rm
det}\int_0^\infty d\omega\, a^{\dag}(\omega)a(\omega) .
\label{eq:Mapprox}
\end{eqnarray}

\subsection{Detection probabilities}
\label{subsec:8C}

The most general state involving $k$ modes can be written as
\begin{equation}
|\psi\rangle = \sum_{n=0} h_n\!:\!\mbox{Pol}_n(a_1^\dag,\ldots, 
a_k^\dag)|0\rangle,
\label{eq:general}
\end{equation}
where $h_n$ is a function of $n$ frequency variables and
$\mbox{Pol}_n(a_1^\dag,\ldots, a_k^\dag)$ is a homogeneous polynomial
of degree $n$ in the $k$ creation operators {\em without any
annihilation operators}.

Let $x$ stand for any of the modes $a_j$. The operator for `detect'
for mode $x$ is $M_1(x)$ and the operator for `no-detect' is obtained
from Eqs.\ (\ref{eq:M1}) and (\ref{eq:unit}) as
\begin{equation}
M_0(x) \defeq  1 - M_1(x) = (1-p_{\rm
  dark}(x))\sum_{n=0}^\infty \,(1-\eta_{\rm det})^n P_{n,+}(x).
\label{eq:M0}
\end{equation}
Any theoretical outcome produced by APD detectors as modeled here is
specified by two lists of modes, a list $\mathbf{J}_0$ for which
APD detectors register `no-detect,' and a list $\mathbf{J}_1$ for
which APD detectors register `detect.'  The operator for `detect' for
modes in the list $\mathbf{J}_1$ and `no-detect' for modes in the
list $\mathbf{J}_0$ is
$\mathbf{M}_1(\mathbf{J}_1)\mathbf{M}_0(\mathbf{J}_0)$, 
where we define
\begin{eqnarray}
\mathbf{M}_0(\mathbf{J}_0) & = & \prod_{x \in  \mathbf{J}_0}M_0(x),
\nonumber \\ 
\mathbf{M}_1(\mathbf{J}_1) & = &  \prod_{x \in  \mathbf{J}_1}M_1(x).
\label{eq:MbfDef}
\end{eqnarray}
The corresponding probability is then 
\begin{equation}
\Pr(\mathbf{J}_0,\mathbf{J}_1) = \langle
\psi|\mathbf{M}_1(\mathbf{J}_1)\mathbf{M}_0(\mathbf{J}_0)|\psi\rangle.
\label{eq:96}
\end{equation}

For $|\psi\rangle$ expressed in the form of Eq.\ (\ref{eq:general})
and the APD model that invokes $M_1$ as defined in Eq.\ (\ref{eq:M1}),
calculating this probability is surprisingly simple.  From Eqs.\
(\ref{eq:M0}) and (\ref{eq:projN}) we obtain the effect of $M_0(a_j)$
on a state vector defined by an integral over frequencies of a
monomial in creation operators
\begin{eqnarray}
M_0(a)\prod_{j=1}^n a^\dag(\omega_j)|0\rangle & = & (1-p_{\rm
dark}(a))(1-\eta_{\rm det})^n \prod_{j=1}^n a^\dag(\omega_j)|0\rangle
\nonumber \\ & = & (1-p_{\rm dark}(a)) \prod_{j=1}^n [(1-\eta_{\rm
det})a^\dag(\omega_j)]|0\rangle.
\label{eq:M0eff}
\end{eqnarray}

{}From this we arrive at the following important

\vspace{3pt}

\noindent\textbf{Proposition}: The effect of an operator $M_0(a)$ for
`no-detection' on a general state $|\psi\rangle$ is to multiply
the state by $(1-p_{\rm dark}(a))$ and to replace every instance of a
creation operator $a^\dag(\omega)$ by $(1-\eta_{\rm
det})a^\dag(\omega_j)$.  Further this holds if creation operators for
modes orthogonal to $a$ enter the polynomial
$\mbox{Pol}_n(a_1^\dag,\ldots, a_k^\dag)$.

\vspace{3pt}

This generalizes to `no-detection' of more modes.

\vspace{3pt}

\noindent\textbf{Proposition}: The effect of a product of creation
operators $M_0(a_1)M_0(a_2)\cdots M_0(a_\ell)$ on a general state 
$|\psi\rangle $ is to multiply the state by
\begin{equation}
\prod_{j=1}^{\ell} (1-p_{\rm dark}(a_j)),
\end{equation}
and to replace every instance of a creation operator
$a_j^\dag(\omega)$, for $j = 1$, \dots, $\ell$, by $(1-\eta_{\rm
det}(a_j))$ $a_j^\dag(\omega_j)$.

\vspace{3pt}

For purposes of calculating probability terms $\langle
\psi|\mathbf{M}_0(\mathbf{L})|\psi\rangle$, where $\mathbf{L}$ is an
arbitrary list of mode names (as in Sec.\ \ref{sec:9}), we can put this
in a symmetric form. Because only like powers of creation and
annihilation operators appear in terms that contribute to
probabilities, we can distribute the factor $(1-\eta_{\rm det}(a_j))$
evenly between the creation and the annihilation operators, as
follows.

\vspace{3pt}

\noindent\textbf{Proposition}: For any list $\mathbf{L}$ of
mode names, one has
\begin{equation}
\langle \psi|\mathbf{M}_0(\mathbf{L})|\psi\rangle = \prod_{x \in
\mathbf{L}} (1-p_{\rm dark}(x))\langle \psi'|\psi'\rangle ,
\label{eq:probEff}
\end{equation}
where $|\psi'\rangle$ is the expression obtained from $|\psi\rangle$
defined in Eq.\ (\ref{eq:general}) by replacing every instance of a
creation operator $a_j^\dag(\omega)$, for $j = 1$, \dots, $\ell$, by
$[1-\eta_{\rm det}(a_j)]^{1/2}a_j^\dag(\omega_j)$, and $\langle
\psi'|$ is obtained from $\langle \psi|$ by the corresponding
replacement of $a(\omega)$ by $[1-\eta_{\rm
det}(a_j)]^{1/2}a_j(\omega_j)$.

\vspace{3pt}

\noindent\textbf{Caution:} This and the preceding two propositions require
that the polynomial in Eq.\ (\ref{eq:general}) contain no annihilation
operators.  

For evaluating $\mathbf{M}_1(\mathbf{J}_1)$, the story is more
complicated.  What makes Proposition (\ref{eq:probEff}) work is that
the substitution of $[1-\eta_{\rm det}(a_j)]^{1/2}a_j^\dag(\omega_j)$
for $a_j^\dag(\omega_j)$ commutes with products, and this does not
hold for the analogous rule for $\mathbf{M}_1$.  From
Eqs.\ (\ref{eq:M0eff}) and (\ref{eq:M0}) we find
\begin{equation}
M_1(a)\prod_{j=1}^n a^\dag(\omega_j)|0\rangle = [1- (1-p_{\rm
dark}) (1-\eta_{\rm
det})^n]\prod_{j=1}^n a^\dag(\omega_j)|0\rangle.
\label{eq:M1eff}
\end{equation}
This proves useful in Sec.\ \ref{sec:9}; however its use is
constrained because it cannot be interchanged with the taking of products
of operators.  For this reason we benefit from the following method of
evaluating $\mathbf{M}_1(\mathbf{L})$ in terms of terms of the form
$\mathbf{M}_0(\mathbf{L}_j)$.  Equations (\ref{eq:M0}) and
(\ref{eq:MbfDef}) imply, for any set $\mathbf{L}$ of mode names,
\begin{equation}
\mathbf{M}_1(\mathbf{L}) = \prod_{x \in \mathbf{L}}[1-M_0(x)].
\end{equation}
With this, we express the product of $M_1$ factors in Eq.\
(\ref{eq:MbfDef}) by
\begin{equation}
\mathbf{M}_1(\mathbf{L}) = \sum_{\mathbf{X}\subset
\mathbf{L}}^{\#(\mathbf{L})}(-1)^{\#(\mathbf{X})}
\mathbf{M}_0(\mathbf{X}),
\end{equation}
where for any set $\mathbf{S}$, $\#(\mathbf{S})$ denotes the number of
elements in $\mathbf{S}$, the sum is over all subsets of $\mathbf{L}$,
including both $\mathbf{L}$ itself and the empty set $\phi$, and we
adopt the convention that
\begin{equation}
\mathbf{M}_0(\phi) = 1.
\end{equation}
Example:
\begin{equation}
\mathbf{M}_1(a_1,b_2) = 1 - \mathbf{M}_0(a_1)- \mathbf{M}_0(b_2) +
\mathbf{M}_0(a_1,b_2).
\end{equation}
Thus we arrive at an equation for the detection operators in Eq.\
(\ref{eq:96}):
\begin{equation}
\mathbf{M}_0(\mathbf{J}_0)\mathbf{M}_1(\mathbf{J}_1) =
\sum_{\mathbf{X} \subset
\mathbf{J}_1}(-1)^{\#(\mathbf{X})}\mathbf{M}_0(\mathbf{J}_0\|\mathbf{X}),
\end{equation}
where $\mathbf{J}_0\|\mathbf{X}$ denotes the concatenation of
the two lists of mode names.  This implies
\begin{eqnarray}
\langle\psi|\mathbf{M}_0(\mathbf{J}_0)\mathbf{M}_1(\mathbf{J}_1)
|\psi\rangle & = & \sum_{\mathbf{X} \subset
  \mathbf{J}_1}(-1)^{\#(\mathbf{X})}
\langle\psi|\mathbf{M}_0(\mathbf{J}_0\|\mathbf{X}) |\psi\rangle
\nonumber \\ & = &
(-1)^{\#(\mathbf{J}_0)}\sum_{\mathbf{X} \subset
    \mathbf{J}_1}(-1)^{\#(\mathbf{J}_0\|\mathbf{X})}
  \langle\psi|\mathbf{M}_0(\mathbf{J}_0\|\mathbf{X}) |\psi\rangle.
\label{eq:formM0}
\end{eqnarray}

\subsection{Effect of time bounds on detection}\label{subsec:8D}
In many applications a detector is gated on only briefly.
Approximating the turn-on and turn-off as perfectly abrupt, we model
the effect of gating the detector on for a duration $T$ centered at a
time $t_g$ by use of Eq.\ (\ref{eq:dt}) with the infinite limits
replaced by finite times $t_g - T/2$ and $t_g + T/2$.  For instance in
Eq.\ (\ref{eq:Mapprox}) we replace $\int_0^\infty d\omega\,
a_j^{\dag}(\omega)a_j(\omega)$ by
\begin{equation}
\int_{t_g-T/2}^{t_g+T/2} dt\, a_{j\pm}^\dag(x,t)a_{j\pm}(x,t)
\label{eq:tdetect}
\end{equation}  
for whichever sense of the $\pm$ sign corresponds to propagation
toward the detector.  Substitution from Eq.\ (\ref{eq:loca}) and
carrying out the time integration yield 
\begin{eqnarray}\lefteqn{
\int_{t_g-T/2}^{t_g+T/2} dt\, a_{j\pm}^\dag(x,t)a_{j\pm}(x,t)
}\quad \nonumber \\
&=& \int_0^\infty\! d\omega \int_0^\infty\!
d\omega'\,a_{j\pm}^\dag(\omega) a_{j\pm}(\omega')
e^{i[(\omega-\omega')t_g
  \mp(k(\omega)-k(\omega'))x]}\,\frac{1}{\pi}\,\frac{\sin[(\omega -
  \omega')T/2]}{\omega- \omega'}.\qquad
\label{eq:gated}
\end{eqnarray}

[* Look at Mandel \cite{mandel}, first at pp.\ 573ff; then pp.\ 691ff.
Check in Mandel, p.~696 (14.2--17), for arguments that dark counts are
in principle unavoidable, due to vacuum fluctuations.]

\subsection{Detection, energy, and photon subspaces}\label{subsec:8E}

Let $\mathcal{H}$ be the vector space of light states generated by
integrals over monomials in creation operators acting on the vacuum
state $|0\rangle$.  Then for a mode $a$, $P_n(a)\mathcal{H}$ is the
subspace of states generated by a weighted integral over $n$ factors,
$\prod_{k=0}^n a^\dag(\omega_k)$, along with any number of factors of
creation operators for modes orthogonal to $a$, all acting on the vacuum
state $|0\rangle$.  Any vector $|v\rangle$ in $\mathcal{H}$ can be
written as a sum over terms in mutually orthogonal subspaces:
\begin{equation} |v\rangle = \sum_{n=0} C_n|v_{n,a}\rangle,
\end{equation}
with $|v_{n,a}\rangle \in P_n(a)\mathcal{H}$ and of unit norm.  Then
we have $\langle v_{n,a}|v_{m,a}\rangle = \delta_{m,n}$. Further,
we have $a(\omega)|v_{0,a}\rangle = 0$ and, for $n > 0$,
$a(\omega)|v_{n,a}\rangle \in P_{n-1}(a)\mathcal{H}$. It follows that
\begin{equation}
(\forall\ \omega,\omega')\quad \langle v| a^\dag(\omega)a(\omega')|v
\rangle
 = \sum_{n=0} \langle v_{n,a}| a^\dag(\omega)a(\omega')| v_{n,a} \rangle.
\end{equation} 
That is, there are no cross terms.  The probability of detection of a
mode $a$, as well as the energy in the mode, are often expressed by
integrals over products $a^\dag(\omega)a(\omega)$ or, per Eqs.\
(\ref{eq:tdetect}) and (\ref{eq:gated}), $a^\dag(\omega)a(\omega')$,
with the result that both the energy for the mode $a$ and the
probability of detection can be expressed as sums with no cross terms
between terms with differing numbers of $a$ photons.  That is, the
expectation energy for energy in the $a$-mode is $\hbar \sum_{n=0} n
|C_n|^2 \omega_n$, where $\omega_n$ is an averaged frequency that is
straightforward to work out.  Similarly, using any of the detection
models discussed above, the probability of detecting the state
$|v\rangle$ by use of a detector involving only the mode $a$ is
$\Pr(\mbox{Detection of }|v\rangle) = \sum_{n=0} |C_n|^2
\Pr(\mbox{Detection of }|v_{n,a}\rangle)$.

\subsection{Preceding the APD detector by a beam-splitter}\label{subsec:8F}
The simplest model of beam splitting (which neglects frequency
dependence, reflection, and mixing of polarizations) expresses an
in-mode $a$ in terms of mutually orthogonal out-modes $b$ and $c$ by
an SU(2) transformation:
\begin{equation} 
a^\dag(\omega) = (\eta_{\rm trans})^{1/2} b^\dag(\omega) + (1-\eta_{\rm
  trans})^{1/2} c^\dag(\omega).\label{eq:b1b2}
\end{equation}

Consider an $n$-photon $a$-mode in-state defined in Eq.\
(\ref{eq:nphot}),
\begin{equation}
|\psi_n\rangle = (n!)^{-1/2} h_n \!:\! a^{\dag n}|0\rangle,
\end{equation}
where, without loss of generality, $h_n$ is symmetric under
permutations of its $n$ arguments, normalized as in Eq.\
(\ref{eq:nnorm}).  The out-state from the SU(2) transformation
expresses the state downstream of a beam-splitter as
\begin{equation}
|\psi_n\rangle = (n!)^{-1/2}\sum_{k=0}^n\left(\begin{array}{c} n
 \\k\end{array}\right)\eta_{\rm trans}^{k/2}(1-\eta_{\rm
 trans})^{(n-k)/2}|\psi_{nk}\rangle,
\end{equation} where we define the unnormalized state vector 
\begin{equation}
|\psi_{nk}\rangle \defeq  (h_n \!:\! b^{\dag
 k}c^{\dag(n-k)})|0\rangle.
\end{equation}

Now calculate the probability of `no-detect' for this state by a
detector of the $b$ mode, described by the operator $M_0(b)$ per the
preceding APD model:
\begin{eqnarray}
\Pr(\mbox{no detect }b)& = &\langle \psi_n|M_0(b)|\psi\rangle\nonumber\\
& = &\frac{1}{n!}\sum_{k=0}^n\left(\begin{array}{c} n
\\k\end{array}\right)^2\eta_{\rm trans}^{k}(1-\eta_{\rm
trans})^{(n-k)}\nonumber\\
& &{}\times\langle 0|  
(h_n^* \!:\! b^kc^{n-k}) M_0(b)(h_n \!:\! b^{\dag k}c^{\dag(n-k)})|0\rangle.
\end{eqnarray}
By Proposition (\ref{eq:probEff}), this becomes
\begin{eqnarray}
 \langle \psi_n|M_0(b)|\psi\rangle & = &  (1-p_{\rm
dark})\frac{1}{n!}\sum_{k=0}^n\left(\begin{array}{c} n
\\k\end{array}\right)^2\eta_{\rm trans}^{k}(1-\eta_{\rm
trans})^{(n-k)}(1-\eta_{\rm det})^k\nonumber\\
& &{}\times \langle 0|
(h_n^* \!:\! b^kc^{n-k})(h_n \!:\! b^{\dag k}c^{\dag(n-k)})|0\rangle.
\end{eqnarray}
{}From Lemmas (\ref{eq:nkrelation}), (\ref{eq:normhn}) of
Appendix~\ref{app:B}, along with the normalization of $h_n$  we have
\begin{equation}
\langle 0|
(h_n^*\!:\!b^kc^{n-k})(h_n\!:\!b^{\dag k}c^{\dag(n-k)})|0\rangle =
k!(n-k)!\ ,
\end{equation}
whence we obtain
\begin{eqnarray}
\langle \psi_n|M_0(b)|\psi\rangle & = &  (1-p_{\rm
dark})\frac{1}{n!}\sum_{k=0}^n\left(\begin{array}{c} n
\\k\end{array}\right)^2\eta_{\rm trans}^{k}(1-\eta_{\rm
trans})^{(n-k)}(1-\eta_{\rm det})^kk!(n-k)!
\nonumber \\ & = & (1-p_{\rm
dark})\sum_{k=0}^n\left(\begin{array}{c} n
\\k\end{array}\right)[\eta_{\rm trans}(1-\eta_{\rm det})]^k(1-\eta_{\rm
trans})^{(n-k)} \nonumber \\ & = & (1-p_{\rm
dark})[\eta_{\rm trans}(1-\eta_{\rm det}) + 1-\eta_{\rm
trans}]^n \nonumber \\
& = & (1-p_{\rm
dark})(1 -\eta_{\rm trans}\eta_{\rm det})^n.
\end{eqnarray}
This is just the probability of `no-detect' for the in-state by 
an $a$-mode detector, modified by replacing $\eta_{\rm det}$ by
the product $\eta_{\rm trans}\eta_{\rm det}$. Thus we have arrived at the

\vspace{3pt}

\noindent\textbf{Proposition}: For the APD model described
above, we consider a beam splitter for
which the free input is a vacuum state, and which passes a
frequency-independent fraction of energy $\eta_{\rm trans}$; then
the probability `no-detect' downstream of the splitter is
obtained from the expression for the probability upstream by
replacing the detector efficiency $\eta_{\rm det}$ by a reduced
efficiency $\eta_{\rm trans}\eta_{\rm det}$.

\vspace{5pt}
\noindent\textbf{Caution}: This proposition holds specifically for
the APD model; there are certainly other models, such as photon
counting, to which it does not apply.

\section{Polarization-entangled light for QKD}\label{sec:9}
Some early models of quantum key distribution 
assumed bi-photon states \cite{ekert91}.  A year ago John Schlafer
asked how the energy of the light pulse---its multi-photon
content---affects the statistics of detection.  As preparation for
Sec.\ \ref{sec:10}, where we start to answer this question, here we
show some of the possible multi-photon states that have to be considered.

\subsection{Bi-photon light states}
The general bi-photon light state is given by Eq.\ (\ref{eq:biPhot}). 
Here we confine ourselves to a special case in which detection
statistics are invariant whenever the same polarization  transform
is performed on both the $a$ and $b$ fibers leaving the source. 
(It will be interesting to check experimentally for this invariance.)
The polarization transformations are the group SU(2), and we refer
to bi-SU(2) invariance as invariance of detection probabilities
under the following frequency-independent transformation, for arbitrary
complex $u,v$ such that $|u|^2 + |v|^2 = 1$,
\begin{equation} 
\left[\begin{array}{c}a_1(\omega) \\a_2(\omega) \end{array}\right]\!\!
  \rightarrow\!\!  \left[\begin{array}{c} u a_1(\omega) + v
 a_2(\omega) \\ -v^* a_1(\omega) + u^* a_2(\omega) \end{array}\right]\!,
\qquad \left[\begin{array}{c}b_1(\tilde{\omega})
 \\b_2(\tilde{\omega}) \end{array}\right]\!\!  \rightarrow\!\!
 \left[\begin{array}{c} u b_1(\tilde{\omega}) + v b_2(\tilde{\omega})
 \\ -v^* b_1(\tilde{\omega}) + u^* b_2(\tilde{\omega})
 \end{array}\right]\!. 
\label{eq:biInv}
\end{equation}

By analyzing the case $u = 0$, $v = 1$, one sees that bi-SU(2)
invariance requires that $h$ in Eq.\ (\ref{eq:biPhot})
satisfy 
\begin{equation}
\left[\begin{array}{ll} 
h_{11}(\omega,\tilde{\omega})\quad & h_{12}(\omega,\tilde{\omega}) \\
h_{21}(\omega,\tilde{\omega}) & h_{22}(\omega,\tilde{\omega})
  \end{array} 
 \right] = g(\omega,\tilde{\omega})\left[
 \begin{array}{rr} 0&\quad 1 \\-1
 &\quad 0 \end{array}\right],
\end{equation}
where $(g,g) = 1$ but $g$ is an otherwise arbitrary function.  Any
such state is a single $a$-photon together with a single
$b$-photon. Sufficiency of this condition to assure bi-photon
invariance is shown by the substitution defined in Eq.\
(\ref{eq:biInv}).

To explore the dependence of detection probabilities on energy, we
need to analyze multi-photon states.  We confine ourselves to an
opening step in this direction by choosing multi-photon states that
have bi-SU(2) invariance of the form
\begin{equation}|\psi\rangle =
\sum_{n=0}C_n|\psi_n\rangle,\vadjust{\kern-5pt}
\end{equation}
where
\begin{eqnarray} |\psi_n\rangle \!& = &\!
 f\!:\!(a_1^\dag b_2^\dag - a_2^\dag b_1^\dag)^n|0\rangle \nonumber \\ 
\!& = &\!\!
\int d\omega_1\,d\tilde{\omega}_1 \cdots d\omega_n\, d\tilde{\omega}_n \,
f(\omega_1,\tilde{\omega}_1,\ldots,\omega_n,\tilde{\omega}_n)
\prod_{j=1}^n\left(a_1^\dag(\omega_j)
b_2^\dag(\tilde{\omega}_j) - a_2^\dag(\omega_j)
b_1^\dag(\tilde{\omega}_j)\right)|0\rangle.\nonumber\\
\end{eqnarray}
Because the factors under the product all commute with one another,
what matters about $f$ is the part that is invariant under
permutations that swap $(\omega_j,\tilde{\omega}_j)$ with
$(\omega_k,\tilde{\omega}_k)$.  We think of a vector $\vec{\omega}_j
=(\omega_k,\tilde{\omega}_k)$ and denote the average of $f$ over
permutations of these vectors by
\begin{equation} 
\mathcal{S}(\vec{\omega}_1,\ldots,\vec{\omega}_n)\,
f(\vec{\omega}_1,\ldots,\vec{\omega}_n) \defeq 
\frac{1}{n!}\sum_{\pi \in S_n} f(\omega_{\pi 1},\tilde{\omega}_{\pi
1},\ldots,\omega_{\pi n},\tilde{\omega}_{\pi n}).
\end{equation}
This is the first of several symmetries that will be seen
to couple polarization entanglement to frequency
entanglement in a way that affects how detection statistics depend on
light energy.

The terms $|\psi_n\rangle$ have the further decomposition
\begin{equation} |\psi_n\rangle = \sum_{m=0}^n|\psi_{nm}\rangle,
\end{equation}
where the unnormalized states $|\psi_{nm}\rangle$ are defined by
\begin{eqnarray}
|\psi_{nm} \rangle & = & (-1)^{n-m}\left(\begin{array}{c}n \\[-5pt] m
 \end{array}\right) \int d\omega_1\,d\tilde{\omega}_1 \cdots
d\omega_n\,d\tilde{\omega}_n \nonumber \\ & &
[\mathcal{S}(\omega_1,\ldots,\omega_m)
\mathcal{S}(\omega_{m+1},\ldots,\omega_n)
\mathcal{S}(\vec{\omega}_1,\ldots,\vec{\omega}_n)\,
f(\omega_1,\tilde{\omega}_1,\ldots,\omega_n,\tilde{\omega}_n)]
\nonumber \\ & &{}\times \left(\prod_{j=1}^m a_1^\dag(\omega_j)
b_2^\dag(\tilde{\omega}_j)\right)\left(\prod_{j= m+1}^n
a_2^\dag(\omega_j) b_1^\dag(\tilde{\omega}_j) \right)|0\rangle,\qquad
\label{eq:gnm}
\end{eqnarray}
with the convention on products defined in Eq.\ (\ref{eq:conven}), and
with the additional symmetry operations that act on $\omega_j$
without acting on $\tilde{\omega}_j$, so that
$S(\omega_1,\ldots,\omega_m)$ is defined for
$m < n$ by
\begin{eqnarray} \lefteqn{ 
S(\omega_1,\ldots,\omega_m)\,f(\omega_1,\tilde{\omega}_1,\ldots,\omega_n,\tilde{\omega}_n)
}\quad \nonumber \\ &=& \frac{1}{m!}\sum_{\pi \in S_{n-m}} f(\omega_{\pi
1},\tilde{\omega}_1,\ldots, \omega_{\pi m},\tilde{\omega}_m,
\omega_{m+1},\tilde{\omega}_{m+1},\ldots, \omega_n,\tilde{\omega}_n).
\end{eqnarray} Similarly, $S(\omega_{m+1},\ldots,\omega_n)$ operates on
the arguments $\omega_j$ for $j = m+1,\ldots,n$.\goodbreak

\noindent\textbf{Remark}:  Although $\mathcal{S}$ as defined in Eq.\
(\ref{eq:Sdef}) is not a quantum operator, it is a linear operator on
a function space, indeed a projection, so that, as an operator,
$\,\mathcal{S}^2 = \mathcal{S}$.  If $h$ is a function with arguments
operated on by $\mathcal{S}$ then $\mathcal{S}h$ is invariant under
the group of permutations over which $\mathcal{S}$ averages.  In
particular, $\mathcal{S}h$ is invariant under the swapping of any two
of the arguments listed in $\mathcal{S}$.  That is the first point.
The second point is that a transposition of $\vec{\omega}_j$ and
$\vec{\omega}_k$ followed by a transposition of $\omega_j$ and
$\omega_k$ is a transposition of $\tilde{\omega}_j$ and
$\tilde{\omega}_k$.  

\subsection{Effect of a beam splitter}
Generalizing on Eq.\ (\ref{eq:b1b2}), if we neglect loss, reflections,
and unwanted polarization couplings, we model the effect of a
non-polarizing beam splitter (or fiber coupler) that splits
say mode $a_j$ into modes $a_{j1}$ and $a_{j2}$ by
\begin{equation} 
a_j^\dag = u a_{j1}^\dag + v a_{j2}^\dag
\label{eq:splitA}
\end{equation} for some complex $u,v$ such that $|u|^2 + |v|^2 = 1$.
To see the effect of this splitting of $a_1$ on the state
$|\psi_{nm}\rangle$, one makes the substitution for $a_1$ in Eq.\
(\ref{eq:gnm}) defined by Eq.\ (\ref{eq:splitA}).  The important
observation is that no further permutation symmetries enter, so the
effect is only to replace
\begin{equation}\prod_{j=1}^m
a_1^\dag(\omega_j)\end{equation}
by 
\begin{equation}
\sum_{k=0}^m\left(\begin{array}{c} m \\ k \end{array} \right) u^k
v^{m-k}\left(\prod_{j=1}^ka_{11}^\dag(\omega_j)\right)\left(
\prod_{j=k+1}^m a_{12}^\dag(\omega_j)\right). 
\end{equation}

\subsection{Effect of polarization rotation}
Unlike beam splitting, the effect of a polarization rotation involves
an added permutation symmetry, and accounting for this symmetry is
essential in calculating the inner products needed to arrive at
probabilities of detection.  This additional symmetry occurs because
polarization rotation mixes, for example, an $a_1$-mode with an
$a_2$-mode.

\newpage
[Part II of this paper contains Section 10, Appendices A through F,
and the References, as outlined in the Table of Contents.]  

\end{document}